\documentclass[aps,prl,reprint,groupedaddress,twocolumn,superscriptaddress]{revtex4-1}
\usepackage{graphicx}     
\usepackage{subfigure}
\usepackage{dcolumn}     
\usepackage{bm}              
\usepackage{hyperref}     
\usepackage{amsmath}   
\usepackage{amssymb}
\usepackage{tabularx}
\usepackage{wasysym}
\usepackage{float}
\usepackage{color}
\usepackage[table]{xcolor}
\usepackage{array}
\usepackage{multirow}
\usepackage{makecell}
\usepackage[version=3]{mhchem} 
\usepackage{float}
\usepackage{lipsum}
\usepackage{ulem}

\usepackage{verbatim}    
 
\makeatletter
\newcommand{\colorcaption}[2][]{%
  \begingroup%
  \renewcommand{\@caption@fignum@sep}{ (color online). }%
  \caption[#1]{#2}%
  \endgroup%
  }
\makeatother

\makeatletter
\newcommand*{\rom}[1]{\expandafter\@slowromancap\romannumeral #1@}
\makeatother

\begin{document}

\title{Functional Renormalization Group Study of Superconductivity \\ in Rhombohedral Trilayer Graphene}

\author{Wei Qin}
\email{weiqin@utexas.edu}
\affiliation{Department of Physics, University of Texas at Austin, Austin TX 78712}
\author{Chunli Huang}
\affiliation{Department of Physics, University of Texas at Austin, Austin TX 78712}
\affiliation{Theoretical Division, T-4, Los Alamos National Laboratory,
Los Alamos, New Mexico 87545, USA}
\author{Tobias Wolf}
\author{Nemin Wei}
\author{Igor Blinov}
\author{Allan H.~MacDonald}
\email{macd@physics.utexas.edu}
\affiliation{Department of Physics, University of Texas at Austin, Austin TX 78712}

\date{\today}

\begin{abstract}
We employ a functional renormalization group approach to ascertain the pairing mechanism and symmetry of the superconducting phase observed in rhombohedral trilayer graphene. 
Superconductivity in this system occurs in a regime of carrier density and displacement field 
with a weakly distorted annular Fermi sea.  We find that repulsive Coulomb interactions can induce 
electron pairing on the Fermi surface by taking advantage of momentum-space structure
associated with the finite width of the Fermi sea annulus. 
The degeneracy between spin-singlet and spin-triplet pairing is lifted by
valley-exchange interactions that strengthen under the RG flow and develop nontrivial momentum-space structure.  
We find that the leading pairing instability is
$d$-wave-like and spin-singlet, and that the theoretical phase diagram
versus carrier density and displacement field agrees qualitatively with experiment.
\end{abstract}

\maketitle

\textit{Introduction--- }Recent experiments have demonstrated that graphene multilayers can exhibit rich strong-correlation physics \cite{Cao:2018aa,Cao:2018ab,Yankowitz:2019aa,Lu:2019aa,Balents:2020aa,Stepanov:2020aa,Saito:2020aa,Wong:2020aa,Zondiner:2020aa,Liu:2021aa,Park:2021aa,Hao:2021aa,Stepanov:2021aa,Saito:2021aa,Cao:2021aa,Choi:2021aa,Choi:2021ab,Jaoui:2022aa}, including broken spin/valley flavor 
symmetries and superconductivity, when the layers are twisted to magic angles that yield extremely flat moir\'e superlattice bands \cite{Bistritzer:2011aa,Khalaf:2019aa,Mora:2019aa,Zhu:2020aa,Lei:2021aa}.  
Twist angle changes during processing make systematic 
studies more difficult, however, and devices inevitably have some twist-angle disorder \cite{Uri:2020aa}. 
For this reason, the recent observation \cite{Zhou:2021aa,Zhou:2021ab} of superconductivity and broken flavor symmetries in untwisted rhombohedral trilayer graphene (RTG) has been a pleasant surprise.
In the regime of displacement field and carrier density where strong correlations have been
observed, the normal state is a two-dimensional hole gas with a distorted annular Fermi surface, which has been precisely characterized using quantum oscillation measurements \cite{Zhou:2021aa,Zhou:2021ab} 
that take advantage of the exceptional sample perfection.  
The strongest superconductivity 
appears as an instability of a flavor-symmetric paramagnetic normal state and has 
an in-plane critical magnetic field that seems most
compatible with spin-singlet pairing \cite{Chandrasekhar:1962aa,Clogston:1962aa}.

Several ideas have already been explored in connection with superconductivity in 
RTG \cite{Chou:2021aa,Chatterjee:2021aa,You:2021aa,Ghazaryan:2021aa,Dong:2021aa,Szabo:2022aa,Cea:2022aa}.  Conventional acoustic-phonon-mediated attraction \cite{Chou:2021aa} can explain superconductivity only if direct Coulomb interactions between electrons 
do not play a significant role.  The experimental observation of a nearly temperature-independent
resistivity up to 20 K suggests that electron-phonon interactions are relatively weak \cite{Zhou:2021aa}, however, arguing against this mechanism.
Inter-valley coherence fluctuation 
mediated pairing was proposed as a possibility \cite{Chatterjee:2021aa,Dong:2021aa,You:2021aa}, 
motivated by the experimental observation that superconductivity is proximal to a phase 
transition to a partially isospin polarized (PIP) phase \cite{Chatterjee:2021aa}. 
However, the observation of a sudden jump in quantum oscillation frequencies 
between superconducting (SC) and PIP phases \cite{Zhou:2021aa} indicates that the phase transition is first order and therefore that  
these critical fluctuations will not be strong.
Since superconductivity in RTG is in the clean limit, the Kohn-Luttinger (KL)
mechanisms \cite{Kohn:1965aa} has also been considered as a possibility \cite{Ghazaryan:2021aa,Cea:2022aa}. It was demonstrated that superconductivity can arise from the combination of 
annular Fermi surfaces and long-range Coulomb repulsion in RTG when a random phase approximation (RPA) is employed \cite{Ghazaryan:2021aa,Raghu:2011aa}.
  
In this Letter, we apply the functional renormalization group (FRG) method to investigate the pairing mechanism and symmetry in RTG.
We start with long-range bare Coulomb interactions, and include a valley-exchange interaction that is $\sim 100$ times weaker than intra- and inter-valley interactions
because the Fermi surfaces are small in size compared to the RTG Brillouin zone.
We find that the valley-exchange interaction is enhanced under the RG flow and develops nontrivial momentum-space structures that cannot be represented by a simple momentum independent inter-valley Hund's coupling \cite{Chatterjee:2021aa,You:2021aa}. The enhanced valley-exchange interaction breaks valley-SU(2) symmetry and lifts the degeneracy between spin-singlet and spin-triplet pairing.  
The renormalized pairing interactions develop features at cross-annulus momentum-transfers 
due to inter-Fermi surface particle-hole fluctuations.  For experimentally relevant 
ratios of the annulus radii, these features 
favor $d$-wave spin-singlet pairing and $p$-wave spin-triplet pairing on the inner and outer Fermi surfaces, respectively.
The competition between pairing and particle-hole channel 
instabilities is sensitive to the Fermi-level density of states as well as the precise annulus shape.

\begin{figure}
 \centering
\includegraphics[width=\columnwidth]{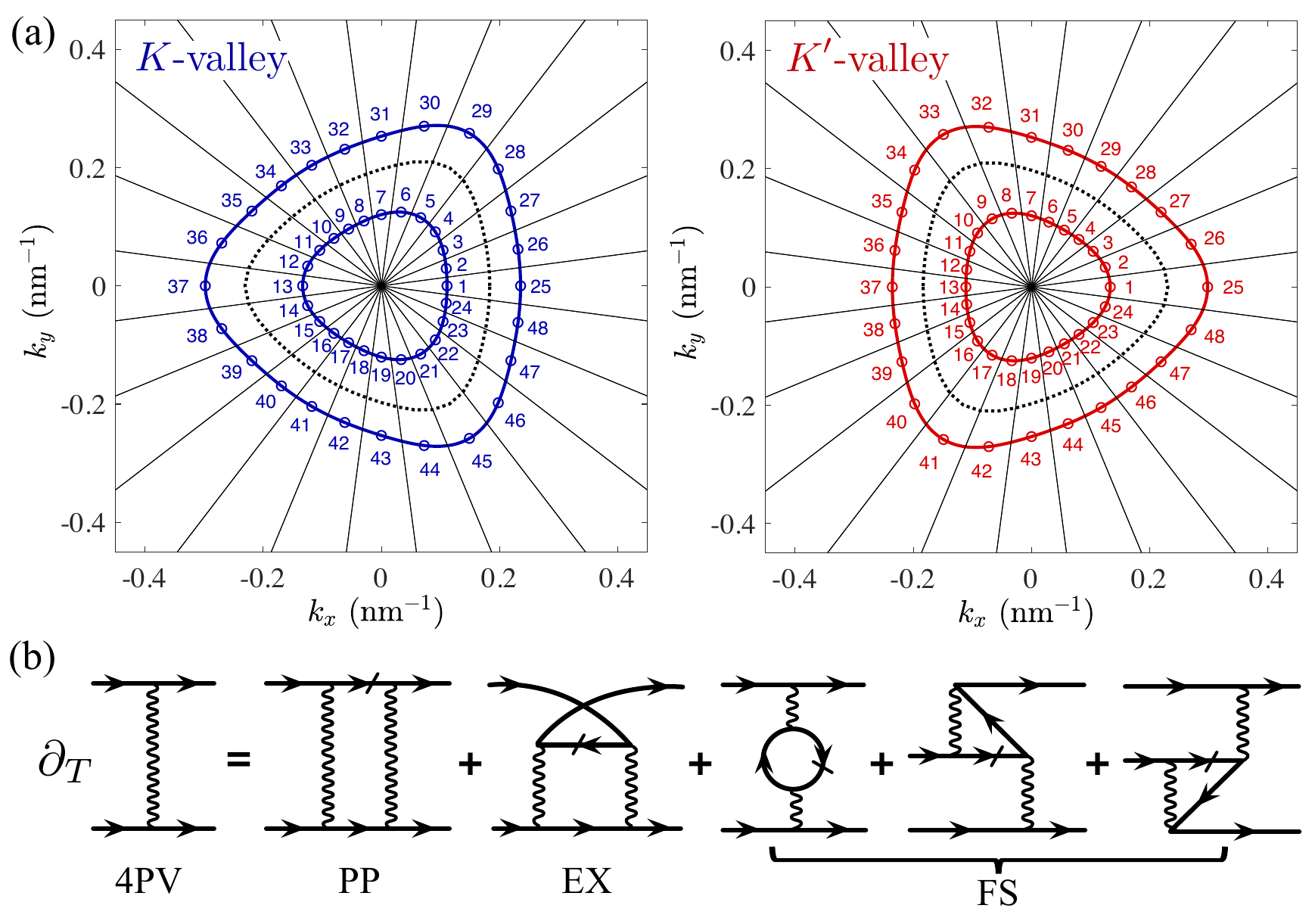}
 \caption{(a) Patching scheme for the annular Fermi surfaces (solid curves). The momentum space around each valley is divided into 48 patches.  The circles on the Fermi surfaces specify
 the wavevectors at which we evaluate 4PVs. 
 The dotted curves identify the wavevectors
 at which energy maxima occur along radial directions, and separate patches that belong to the inner and outer Fermi surfaces in our FRG calculations. (b) Diagrammatic representation of the one-loop RG flow equation for
 4PVs, including particle-particle (PP), exchange (EX) and forward scattering (FS) 
 contributions. Here the temperature $T$ is the RG flow time, and the slashes on the propagators denote temperature derivatives \cite{Honerkamp:2001aa}.}
 \label{fig:figure1}
\end{figure}

\textit{Model---}
We employ the well established six-band continuum model detailed in the 
Supplementary Information (SI) \cite{SM} that accounts for long-range Coulomb interactions, gate and dielectric screening, and all pertinent details of the low-energy electronic structure.  In the following calculations, we choose the displacement field-induced electrostatic potential $\Delta_d$ = 30 meV and carrier density $n_e =-1.75\times 10^{12}$ cm$^{-2}$ as representative \cite{SM}. 
The corresponding annular Fermi surfaces are illustrated in Fig.~{\ref{fig:figure1}}(a).
To capture the competition between superconductivity, inter-valley coherence, and spin-polarized half metals in RTG, we employ the temperature-flow FRG scheme because it properly accounts for small-$\bm{q}$ particle-hole fluctuations \cite{Metzner:1998aa,Halboth:2000aa,Honerkamp:2001aa}. 
The technical details of the present study are similar to earlier works \cite{Shankar:1994aa,Honerkamp:2001ab,Honerkamp:2001aa,Metzner:2012aa}, especially those in multi-orbital systems with more than one Fermi surface \cite{Wang:2009aa,Thomale:2011aa,Platt:2013aa}. 
To implement numerical FRG calculations, we set the initial temperature to $T_0 \sim 11600$ K \cite{SM} and employ the $N$-patch scheme \cite{Honerkamp:2001aa,Sante:2022aa} illustrated in Fig.~\ref{fig:figure1}(a).
  
\begin{figure}
 \centering
\includegraphics[width=\columnwidth]{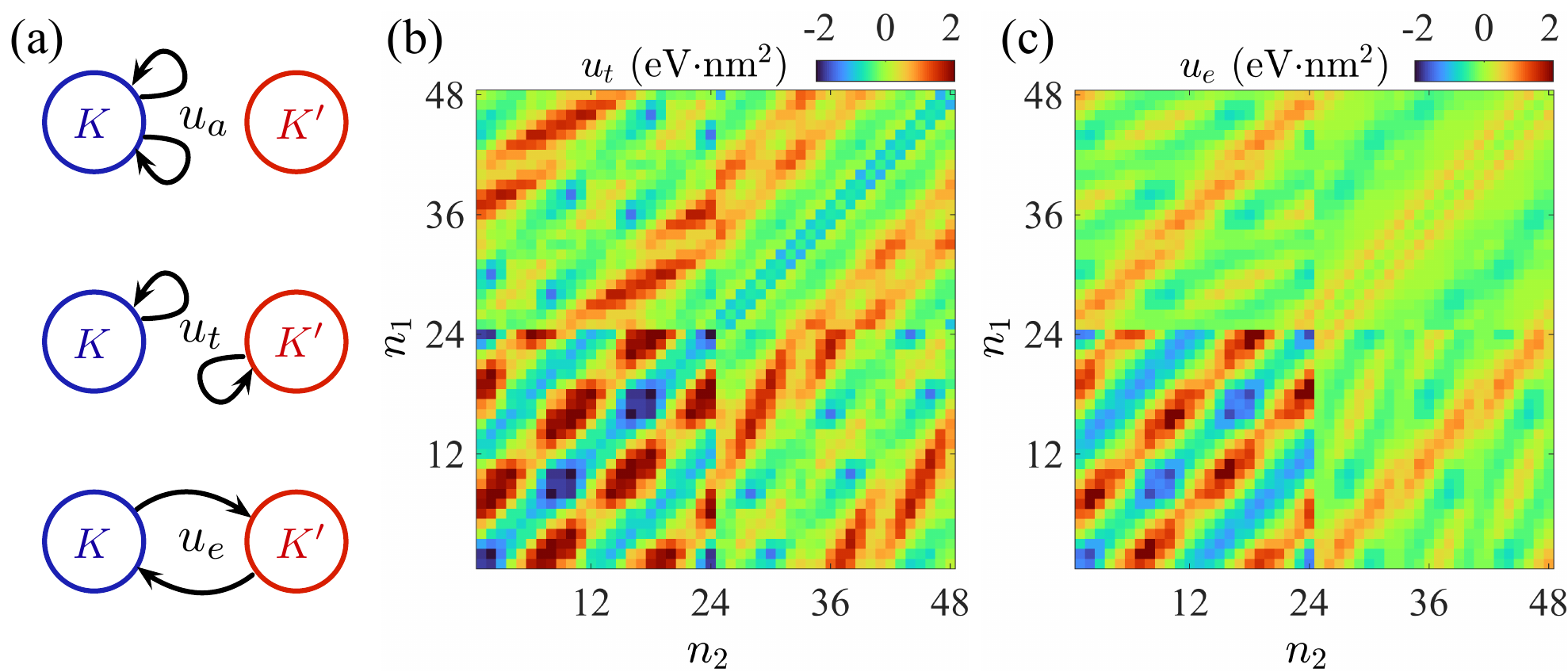}
 \caption{(a) Schematic diagrams of $u_{a,t,e}$.  (b)-(c) Renormalized pairing interactions at  temperature $T=70$ mK, including (b) $u_t(n_1,\bar{n}_1;\bar{n}_2)$ and (c) $u_e(n_1,\bar{n}_1;n_2)$. The bottom-left and top-right 24$\times$24 blocks are scatterings on the inner and outer Fermi surfaces, respectively. The remaining blocks are scatterings between the inner and outer Fermi surfaces. 
 }
 \label{fig:figure2}
\end{figure}

The one-loop FRG flow equations for 4-point vertices (4PVs) with spin-SU(2) symmetry are depicted diagrammatically in Fig.~\ref{fig:figure1}(b). These diagrams are similar to KL-mechanism diagrams \cite{Kohn:1965aa}, 
previously employed to study pairing instabilities in graphene-based systems \cite{Gonzalez:2008aa,Gonzalez:2013aa,Kagan2014aa,Kagan2014ab,Kagan2015aa,Goodwin2019aa,Gonzalez:2019aa,Phong:2021aa,Ghazaryan:2021aa,Cea:2022aa}. The essential difference between the KL diagram and the FRG flow equation is that the former focuses only on the irreducible pairing vertex while the latter retains all 4PVs on an equal footing, making it possible to explore competing orders. Moreover, particle-hole fluctuations that do not develop instabilities can upon reducing temperature still give progressively larger contributions to the RG flow of the pairing vertex, leading to an enhancement of the pairing instability.
Appealing to standard rescaling and power counting arguments \cite{Shankar:1994aa,Metzner:2012aa}, 
we focus on the zero-frequency 4PVs with momenta on the Fermi surface.  We therefore choose the wavevectors marked by circles in Fig.~\ref{fig:figure1}(a) to represent each patch and approximate the 4PVs by
$u(\bm{k}_1,\bm{k}_2;\bm{k}_3) = u(n_1,n_2;n_3)$ for all wavevectors $\bm{k}_{i}$ in the same patch $n_i$ \cite{caveat1}. 
The fourth wavevector is determined by momentum conservation. 
As shown in Fig.~\ref{fig:figure2}(a), the 4PVs can be classified as intra-valley ($u_a$), 
inter-valley ($u_t$), or valley-exchange ($u_e$) using the definitions 
\begin{equation}
\begin{aligned}
u_a(n_1,n_2;n_3) &= u(K n_1,Kn_2;Kn_3),  \\
u_t(n_1,n_2;n_3) &= u(Kn_1,K'n_2;K'n_3), \\
u_e(n_1,n_2;n_3) &= u(Kn_1,K'n_2;Kn_3),
\label{eq:classify}
\end{aligned}
\end{equation}
where the patch indices are associated with valleys.
For example, $n_2$ in $u_a$ and $u_t$ are on $K$- and $K'$-valley Fermi surfaces, respectively. 
Time-reversal symmetry requires $u(\tau_1n_1,\tau_2n_2;\tau_3n_3) =  u^*(\bar{\tau}_1\bar{n}_1,\bar{\tau}_2\bar{n}_2;\bar{\tau}_3\bar{n}_3)$,
where $\bar{n}$ denotes the opposite patch number of $n$ on the same Fermi surface. 
Using these conventions, the valley index can be removed from the 
FRG flow equations (see SI \cite{SM}).

\begin{figure*}
 \centering
\includegraphics[width=1.99\columnwidth]{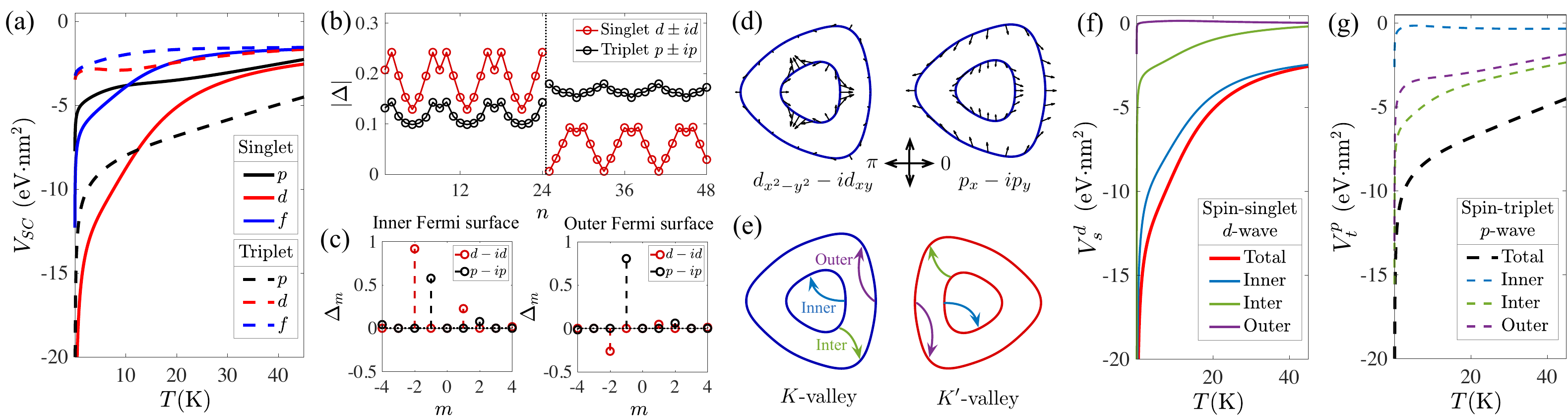}
 \caption{(a) Temperature flows of pairing interaction in channels distinguished by gap function symmetries. (b) Magnitudes of spin-singlet $d \pm id$ and spin-triplet $p \pm ip$
 gap functions on the Fermi surfaces. (c) Angular momentum amplitudes $\Delta_m$ 
 of the gap function on the inner (left) and outer (right) Fermi surfaces.
 (b)-(c) are calculated at $T=70$ mK.
 (d) Spin-singlet $d-id$ and spin-triplet $p-ip$ gap functions 
 on the $K$-valley Fermi surfaces, where the length and direction of arrows 
 specify gap amplitude and phase. (e) Schematic diagram of the inner-, inter-, and outer-Fermi surface scatterings that constitute the total pairing interaction $u_t(n_1,\bar{n}_1;\bar{n}_2,n_2)$. Similar definitions are used for $u_e(n_1,\bar{n}_1;n_2,\bar{n}_2)$.
 (f)-(g) Contributions from the three types of scatterings illustrated in (e) to pairing interactions in (f) spin-singlet $d$-wave and (g) spin-triplet $p$-wave channels.}
 \label{fig:figure3}
\end{figure*}

\textit{Renormalized pairing interaction---} The initial values of the 4PVs are obtained by patch averaging bare Coulomb interactions using band eigenstates to calculate wave function overlap factors 
(see Sec. IV in SI \cite{SM}).
For opposite-valley electron pairing, the relevant 4PVs are $u_t(n_1,\bar{n}_1;\bar{n}_2)$ and $u_e(n_1,\bar{n}_1;n_2)$, where the outgoing momenta belong to $K$ and $K'$ valleys. 
The initial values of $u_e$ are much weaker (less than $1\%$) than those of $u_t$ because the magnitude of the inter-valley momentum-transfer is much larger than the typical size of the annular Fermi surfaces in RTG \cite{SM}.
As shown in Figs.~\ref{fig:figure2}(b) and (c), $u_t$ and $u_e$ flow to comparable magnitudes as temperature reduced to $T=70$ mK. 
The effective pairing interactions are
\begin{equation}
V_{s,t} = u_t(n_1,\bar{n}_1;\bar{n}_2,n_2) \pm u_e(n_1,\bar{n}_1;n_2,\bar{n}_2),
\label{eq:pairing_interaction}
\end{equation}
where the upper (lower) sign is for spin-singlet (spin-triplet) pairing, and the fourth patch index is restored based on momentum conservation.
The results shown in Figs.~\ref{fig:figure2}(b) and (c) indicate 
that the inner Fermi surface prefers spin-singlet pairing because $u_t$ and $u_e$ share similar momentum-space structure, resulting in $|V_{s}| > |V_{t}|$. In contrast, scattering across the annular Fermi surfaces and on the outer Fermi surface prefers spin-triplet pairing because $u_t$ and $u_e$ possess opposite momentum-space structures, leading to $|V_{s}| < |V_{t}|$.

\textit{Pairing symmetry---} The gap function at $T_c$ is
specified by the solution of the linearized gap equation. For the present study, we find that the spin-singlet and spin-triplet gap functions are mainly determined by the momentum-space structures of $V_{s,t}$ and can be well approximated by the eigenvectors of the largest magnitude negative eigenvalues of $V_{s,t}$ (see details in the SI \cite{SM}). Figure~\ref{fig:figure3}(a) plots several of the lowest eigenvalues of $V_{s,t}$ as a function of $T$. We find that the strongest spin-singlet and spin-triplet channels
are respectively $d$-wave-like and $p$-wave-like, and that both channels have doubly degenerate eigenvectors.
Figure~\ref{fig:figure3}(b) plots the magnitudes of the chiral $d\pm id$ and $p \pm ip$ combinations of these eigenvectors, which are stablized below the critical temperatures \cite{SM}. Note that the leading pairing instability is spin-singlet $d$-wave-like because it possesses the lowest eigenvalue as shown in Fig.~\ref{fig:figure3}(a). The deep minima in the gap function on the outer Fermi surface in Fig.~\ref{fig:figure3}(b) are due to strong mixing between different angular momentum channels. The terminology used to distinguish $p$- and 
$d$-wave pairings is justified 
by the Fourier coefficients of both gap functions given in Fig.~\ref{fig:figure3}(c), where one angular momentum is always dominant. Figure.~\ref{fig:figure3}(d) illustrates
the $d-id$ and $p-ip$ gap structures, which are largest on the inner Fermi surface for spin-singlet $d$-wave pairing and are comparable in magnitude on the inner and outer Fermi surfaces for spin-triplet $p$-wave pairing.

Figure~\ref{fig:figure3}(f)-(g) decompose the spin-singlet $d$-wave ($V_{s}^{d}$) and spin-triplet $p$-wave ($V_{t}^{p}$) pairing interactions into contributions from inner-, inter-, and outer-Fermi surface scatterings illustrated in Fig.~\ref{fig:figure3}(e) (details in SI \cite{SM}). We find that $V_{s}^{d}$ is dominated 
by scattering on the inner Fermi surfaces while the strongest contribution to $V_{t}^{p}$ is the inter-Fermi surface scattering, which explains the comparable pairing amplitudes on the annular Fermi surfaces in the spin-triplet $p$-wave channel. The ultimate pairing symmetry depends on the competition between all three types of scattering, which depends in turn on the Fermi surface shape as we show later.

\begin{figure}
 \centering
\includegraphics[width=\columnwidth]{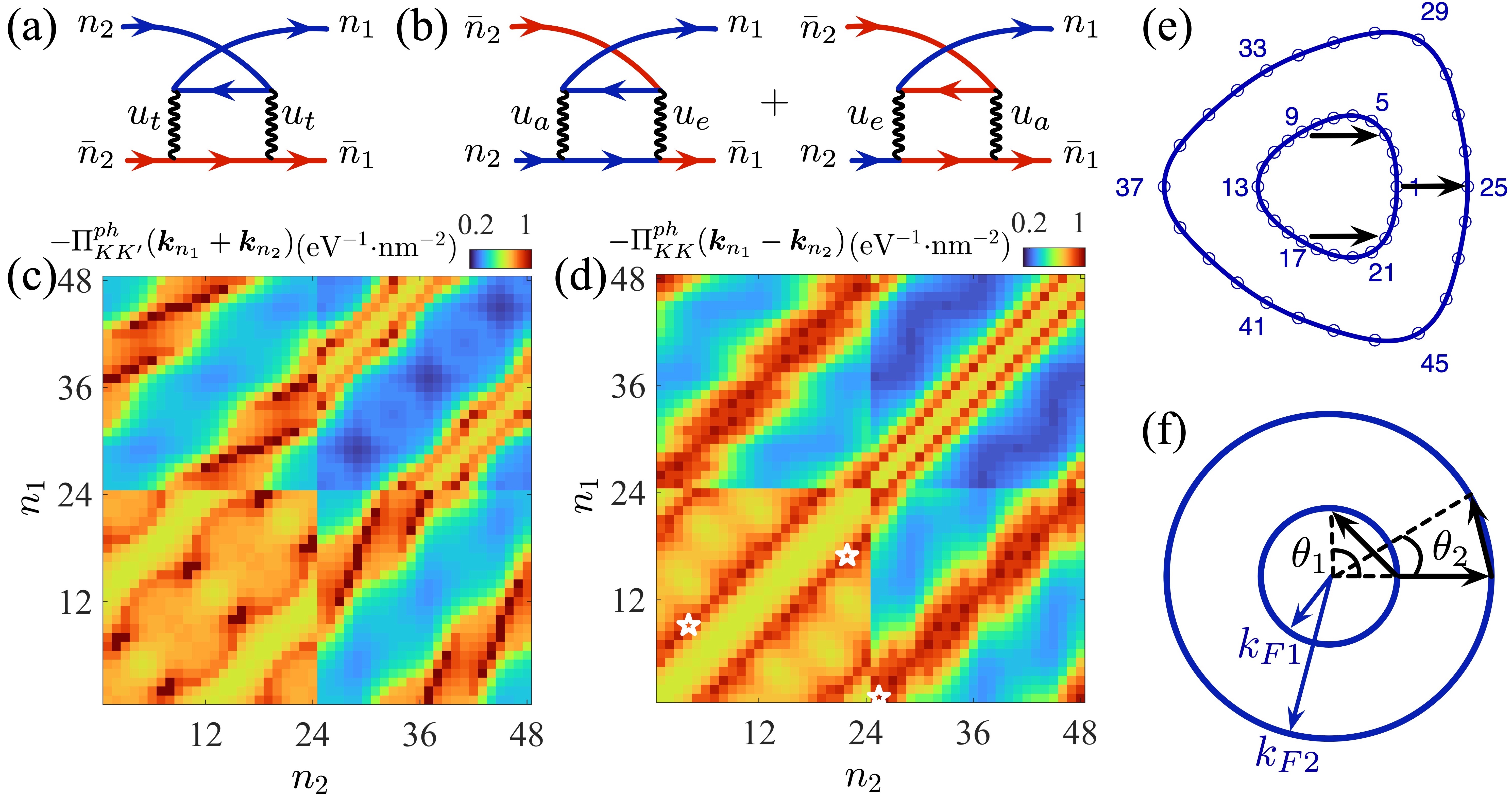}
 \caption{(a)-(b) EX diagram contributions to $u_t(n_1,\bar{n}_1;\bar{n}_2,n_2)$ and $u_e(n_1,\bar{n}_1;n_2,\bar{n}_2)$, where blue (red) solid lines denote $K$($K'$)-valley propagators. (c)-(d) Particle-hole susceptibilities associated with (a)-(b), calculated at $T=70$ mK. (e) Three equivalent inter-Fermi-surface nesting wavevectors $\bm{k}_{25}-\bm{k}_{1} \approx \bm{k}_{4}-\bm{k}_{9} \approx \bm{k}_{22}-\bm{k}_{17} $, at which $\Pi^{ph}_{KK}$ peaked: $(n_1,n_2 ) = (1,25)$, $(9,4)$, and $(17,22)$ marked by stars in (d). 
 (f) Circular annular Fermi surfaces: $k_{F1,2}$ are the two Fermi wavevectors. $\theta_{1,2}$ are the angles spanned by $k_{F2}-k_{F1}$ on the inner and outer Fermi surfaces.}
 \label{fig:figure4}
\end{figure}

\textit{Pairing mechanism---} Figure~\ref{fig:figure4}(a) and (b) depicts the EX diagram contributions to $u_t$ and $u_e$, and the associated inter-valley ($\Pi_{KK'}^{ph}$) and intra-valley ($\Pi_{KK}^{ph}$) particle-hole susceptibilities are plotted in Figs.~\ref{fig:figure4}(c) and (d). By comparing
Figs.~\ref{fig:figure4}(c)-(d) with Figs.~\ref{fig:figure2}(b)-(c), we find
that the momentum-space structure of these susceptibilities is responsible for the momentum-space structure developed in the pairing interactions under FRG, which in turn control the pairing symmetry discussed earlier.
For RTG, the momentum-space structure of $\Pi_{KK'}^{ph}$ and $\Pi_{KK}^{ph}$ arises
primarily from inter-Fermi surface nesting, as illustrated in Figs.~\ref{fig:figure4}(d) and (e). 
In fact, the high-temperature initial stage of the RG flow is dominated by FS processes that screen the long-range Coulomb interaction. The EX enhancement becomes important in the intermediate temperature regime, giving rise to the momentum-space structure of the paring interactions, which is then
amplified by the PP diagram upon further reducing temperature \cite{SM}. Therefore, we argue that the KL-like finite angular momentum pairing we have found stems from particle-hole fluctuations that are enhanced by inter-Fermi surface nesting.

It is instructive to analyze the pairing symmetry using the simplified circular Fermi surface model shown in Fig.~\ref{fig:figure4}(f) for which the electron energy spectrum is valley independent \cite{SM}. Therefore, $\Pi^{ph}_{KK} (\bm{k}_{n_1}-\bm{k}_{n_2}) = \Pi^{ph}_{KK} (\theta)$ and $\Pi^{ph}_{KK'} (\bm{k}_{n_1}+\bm{k}_{n_2})= \Pi^{ph}_{KK} (\pi+\theta)$, 
where $\theta$ denotes the angle spanned by $\bm{k}_{n_1}-\bm{k}_{n_2}$ on the Fermi surfaces. The $\pi$-phase difference between $\Pi^{ph}_{KK} $ and $\Pi^{ph}_{KK'} $ is distorted by trigonal warping in Figs.~\ref{fig:figure4}(c) and (d). Performing Fourier series expansion in $\theta$,
\begin{equation}
-\Pi^{ph}_{KK} (\theta) = \sum_m A_m \cos(m \theta).
\label{eq:PIFourier}
\end{equation}
Since the momentum-space structures of the pairing interactions are primarily determined by those of the particle-hole susceptibilities, we approximate $\delta u_t (\theta)   \propto -\langle u_t \rangle^2 \Pi^{ph}_{KK'}(\theta)  $ and $\delta u_e(\theta)   \propto - \langle u_a \rangle \langle u_e \rangle \Pi^{ph}_{KK}(\theta)$, where $\langle u_{a,t,e}\rangle$ denote their averaged values over $\theta$  \cite{SM}. Combining with Eqs.~(\ref{eq:pairing_interaction})-(\ref{eq:PIFourier}), the singlet and triplet pairing interactions in the $m$-wave channel
$
V_{s,t}^{m} \propto \left[(-1)^m \pm 1\right]A_m.
$
As shown in Fig.~\ref{fig:figure4}(f), inter-Fermi surface nesting enhances $-\Pi^{ph}_{KK} (\theta)$ around $\theta_{1}$ and $\theta_{2}$ on the inner and outer Fermi surfaces.  
For this study, $\theta_{1} \sim \pi/2$, leading to
$-A_2>|A_1|$, and hence preferring spin-singlet $d$-wave ($m=2$) pairing on the inner Fermi surface.
In contrast, $\theta_{2}<\pi/6$, which results in $A_1>|A_2| $ and prefers spin-triplet $p$-wave ($m$=1) pairing.
Inter-Fermi surface scattering always prefers spin-triplet $p$-wave pairing because $-\Pi^{ph}_{KK} (\theta)$ exhibits a maximum at $\theta = 0$, as indicated in Fig.~\ref{fig:figure4} (d).
Overall, the competition between $d$-wave singlet and $p$-wave triplet pairings in RTG is 
sensitive to the shape of the annular Fermi surfaces.  

\begin{table}
\centering
\caption{Matrix structures and symmetries under time-reversal $\mathcal{T}$ and inversion $C_2$ operations of the order parameters for Pomeranchuk instability (PI), valley polarization (VP), ferromagnetism/anti-ferromagnetism (FM/AFM), and inter-valley charge/spin coherence (IVC/IVS). Here FM and AFM are distinguished by identical and opposite spin polarizations in two valleys, $s_{x,y,z}$ and $\tau_{x,y,z}$ are Pauli matrices in spin and valley subspaces, and $\mathcal{K}$ denotes complex conjugate.}
\label{tab:table1} 
\begin{tabular}{ c | c |c |c |c |c | c  }
\hline
\hline
 Particle-hole instability & PI & VP & FM & AFM & IVC & IVS \\
 Order parameter & $s_0\tau_0$  &  $s_0\tau_z$ & $s_z\tau_0$  & $s_z\tau_z$  & $s_0\tau_{x,y}$  & $s_z\tau_{x,y}$ \\
 \hline
 $\mathcal{T} = is_y\tau_x \mathcal{K}$  & \checkmark & $\times$ & $\times$ & \checkmark & \checkmark & $\times$ \\
 $C_2 = s_0\tau_x$ & \checkmark & $\times$ & \checkmark & $\times$ & $\times$ & $\times$ \\
 \hline
 \hline
\end{tabular}
\end{table}

\begin{figure}
 \centering
\includegraphics[width=\columnwidth]{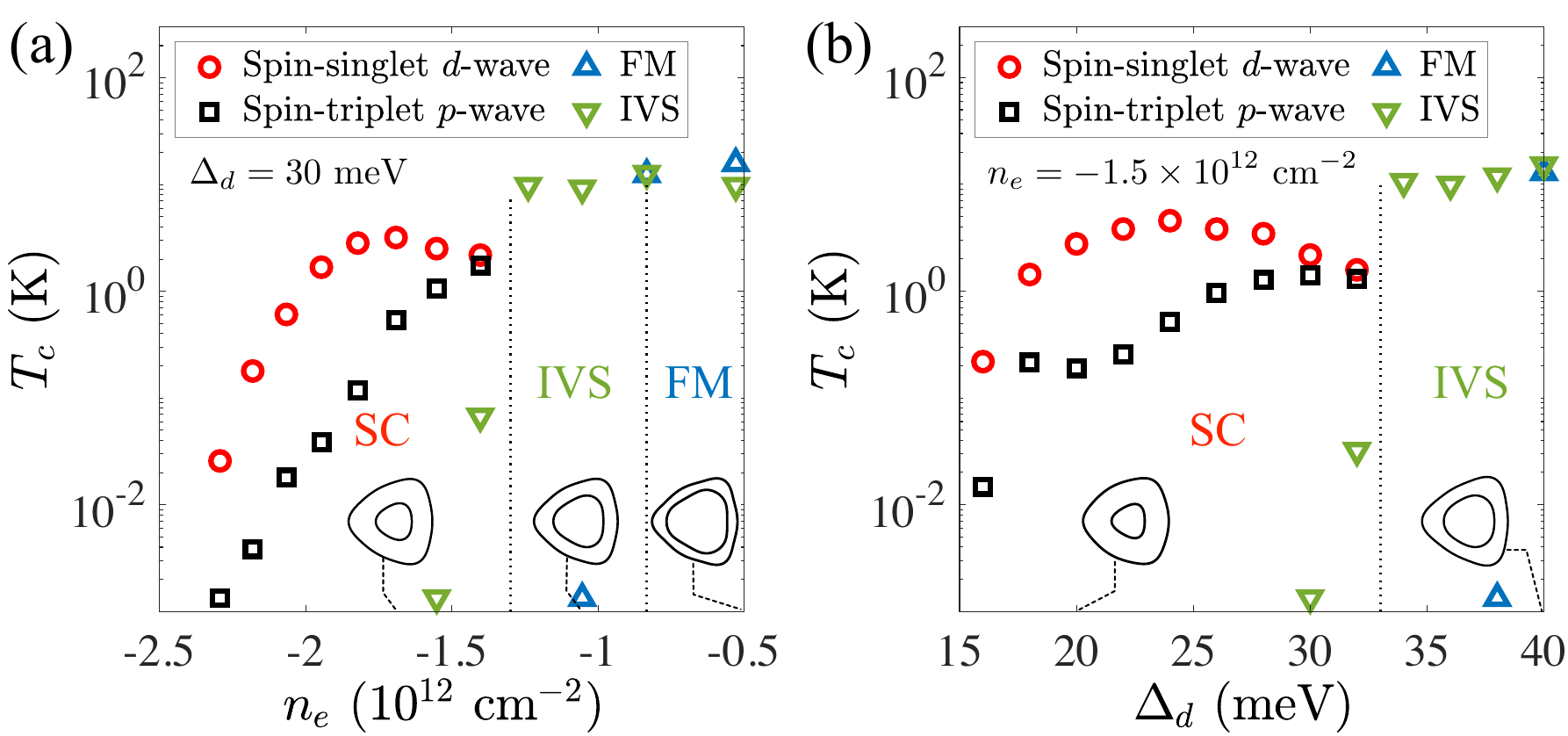}
 \caption{Phase diagrams as function of (a) $n_e$ and (b) $\Delta_d$. The inserts depict the annular Fermi surfaces for several typical values of $n_e$ and $\Delta_d$ that are associated with the SC, IVS and FM states. These results are obtained by choosing dielectric constant $\epsilon=4$ and gate-sample distance $d_s = 40$ nm.}
 \label{fig:figure5}
\end{figure}

\textit{Phase diagram---} 
Quantitative prediction of superconducting critical temperature $T_c$
from FRG calculations stands as a challenging issue [44,47]. 
Here we provide crude $T_c$ estimates by solving linearized gap 
equations with an effective interaction (see details in Sec. VI of SI [39]). 
For the electron pairing channel, the linearized gap equation is
\begin{equation}
\Delta_{s,t}(n) = -\sum_{n_1} \Pi_{KK}^{pp}(n_1) V_{s,t}(n_1,\bar{n}_1;\bar{n},n) \Delta_{s,t}(n_1),
\end{equation}
where $\Delta_{s,t}$ are spin-singlet and spin-triplet order parameters, $\Pi_{KK}^{pp}(n_1) $ denotes the particle-particle susceptibility, and the renormalized pairing interactions $V_{s,t}$ 
are given by Eq.~(\ref{eq:pairing_interaction}). 
Since both $\Pi_{KK}^{pp}$ and $V_{s,t}$ are temperature dependent, $T_c$ is estimated for each channel as
the temperature at which the corresponding eigenvalue of $-\Pi_{KK}^{pp} V_{s,t}$ equals 1.

We have also explored the competition between pairing and particle-hole channel instabilities. 
Table.~\ref{tab:table1} summarizes several typical particle-hole channel instabilities (see details in Sec.~VI of SI \cite{SM}). 
As shown in Fig.~\ref{fig:figure5}(a), for large hole densities that are associated with relatively thick annular Fermi seas, spin-singlet $d$-wave pairing is the leading instability. 
The corresponding $T_c$ exhibits a dome-like behavior {\it vs.} hole density that is  
qualitatively consistent with experimental observations \cite{Zhou:2021aa}. 
We note that experimental signatures of a SC state do appear to be present in the longitudinal resistivity at hole densities that
exceed the region that can currently be identified as in the SC dome \cite{Zhou:2021aa} 
The data seem to be consistent with the notion that superconductivity survives to higher hole densities, as in our calculations, albeit with substantially decreased $T_c$. By decreasing the hole density and moving the Fermi level toward the VHS (see Fig.~\ref{fig:figure5}(a)), the PIP state seen in experiment appears consistent with an IVS state emerging first and then being replaced by the FM state.  In our calculations, these two states compete closely when the Fermi level of the system approaches the VHS. Since the averaged value of $u_e$ is positive, IVS and FM always dominate over IVC and AFM (see SI \cite{SM}). Similar behavior is revealed in the phase diagram {\it vs.} $\Delta_d$ for a given $n_e$ in Fig.~\ref{fig:figure5}(b). The qualitative features of these phase diagrams are therefore in good agreement with experimental observations \cite{Zhou:2021aa}. 

\textit{Discussion---} Numerical estimates of 
superconducting $T_c's$ depend on model parameters, such as the dielectric constant $\epsilon$ (see SI \cite{SM}).  One general trend is that stronger Coulomb interaction results in higher $T_c$, contradictory to the acoustic-phonon-mediated superconductivity \cite{Chou:2021aa}. The experimental phase diagram \cite{Zhou:2021aa} shows that the PIP (IVS) state emerges at a larger hole density than in our theoretical estimation. If the stability region of 
the IVS state is expanded to agree with experiment, it intervenes before the peak of the superconducting dome is reached. This shift would avoid the region in the phase diagram where the $p$-wave spin-triplet pairing state becomes nearly degenerate with or even dominates over the $d$-wave spin-singlet pairing \cite{SM}. In fact, an accurate determination of the phase boundary is obviously very demanding and beyond the scope of the present FRG study. The sudden jumps of the $T_c's$ for the IVS and FM states shown in Fig.~\ref{fig:figure5} arise from the fact that the particle-hole susceptibilities possess maxima at finite temperatures due to proximity to the VHS \cite{SM}. Such a jump in $T_c$ for the IVS state is consistent with experimental observation of the jump in quantum oscillation frequency between SC and PIP phases \cite{Zhou:2021aa}. We have explored the effects of changing other model parameters \cite{SM}. The resulting phase diagrams share similar qualitative features with Fig.~\ref{fig:figure5}. 

The present study explains the spin-singlet SC dome observed in RTG
in terms of inter-Fermi-surface nesting that leads to
$d$-wave singlet pairing. 
We find that the Fermi surface geometry plays
the key role in determining pairing symmetry and predict that the strongest
$d$-wave spin-singlet pairing emerges when the ratio of the enclosed areas of the inner
and outer Fermi surfaces is $r = S_{in}/S_{out} \approx 0.176$,
close to the value of $0.2$ measured near the SC dome \cite{Zhou:2021aa,SM}.
Larger values of $r$ obtained by increasing $\Delta_d$ or $n_e$ lead to first-order spin/valley
ferromagnetic phase transitions that reconstruct the Fermi surface \cite{Huang:2022aa}, removing the annulus and 
preempting the pairing instability.  

\begin{acknowledgments}
\textit{Acknowledgment ---} 
The authors are grateful for helpful interactions with Andrea F. Young, Haoxin Zhou and Jihang Zhu. This work was supported by the U.S. Department of Energy, Office of Science, Basic Energy Sciences, under Award DE-SC0022106. The work done at LANL was carried out under the auspices of the US DOE NNSA under Contract No.~89233218CNA000001 through the LDRD Program. T.M.R.W.~is grateful for the financial support from the Swiss National Science Foundation (Postdoc.Mobility Grant No.~203152).
\end{acknowledgments}


\begin{thebibliography}{99}
\bibitem{caveat1}In our convention $u(\bm{k}_1,\bm{k}_2;\bm{k}_3)$ represents the Coulomb matrix elements where the incoming momenta $\bm{k}_3-\bm{k}_1-\bm{k}_2$ and $\bm{k}_3$ scatter respectively into outgoing momenta $\bm{k}_1$ and $\bm{k}_2$.
\end{thebibliography}
\end{document}


\title{Supplemental Information: Functional Renormalization Group Study of Superconductivity in Rhombohedral Trilayer Graphene}

\author{Wei Qin}
\email{weiqin@utexas.edu}
\affiliation{Department of Physics, University of Texas at Austin, Austin TX 78712}
\author{Chunli Huang}
\affiliation{Department of Physics, University of Texas at Austin, Austin TX 78712}
\affiliation{Theoretical Division, T-4, Los Alamos National Laboratory,
Los Alamos, New Mexico 87545, USA}
\author{Tobias Wolf}
\author{Nemin Wei}
\author{Igor Blinov}
\author{Allan H.~MacDonald}
\email{macd@physics.utexas.edu}
\affiliation{Department of Physics, University of Texas at Austin, Austin TX 78712}

\date{\today}

\maketitle

\section{Band structure model}
We employ the $\pi$-band continuum model \cite{Zhang:2010aa} to describe the low-energy physics 
of rhombohedral trilayer graphene (RTG) which is controlled by states near the triangular lattice Brillouin zone (BZ) corners.  The valley-projected band Hamiltonian 
\begin{equation}
h(\bm{k}) = 
\begin{pmatrix}
\Delta_d+\delta_1  + \delta_2 -\mu & \gamma_2/2 & v_0 \pi^{\dagger} & v_4 \pi^{\dagger} & v_3 \pi & 0 \\
  \gamma_2/2  &   -\Delta_d +\delta_1 + \delta_2  -\mu& 0 &  v_3 \pi^{\dagger} & v_4 \pi & v_0 \pi  \\
  v_0 \pi & 0 & \Delta_d + \delta_1 -\mu & \gamma_1 & v_4 \pi^{\dagger} & 0 \\
 v_4 \pi &  v_3 \pi & \gamma_1 & -2 \delta_1 -\mu  & v_0 \pi^{\dagger} & v_4 \pi^{\dagger} \\
 v_3 \pi^{\dagger}  & v_4 \pi^{\dagger} & v_4 \pi & v_0 \pi & -2 \delta_1  -\mu&\gamma_1 \\
 0 &v_0 \pi^{\dagger} & 0 & v_4 \pi & \gamma_1 &  -\Delta_d+\delta_1 -\mu
 \end{pmatrix},
 \label{eq:bandmodel}
\end{equation}
where the $\pi$-orbital spinors $\psi_{\bm{k}} = (\psi_{1A\bm{k}},\psi_{3B\bm{k}},\psi_{1B\bm{k}},\psi_{2A\bm{k}},\psi_{2B\bm{k}},\psi_{3A\bm{k}})^{T}$, $\psi_{l\sigma \bm{k}}$ annihilates an electron with wavevector $\bm{k}$ on
sublattice $\sigma=A,B$ and layer $l=1,2,3$, $\pi = \tau k_x+ik_y$ with valley index $\tau = \pm$ 
distinguishing $K$ and $K'$ valleys, and $\mu$ is chemical potential. The Dirac velocity $v_i = \sqrt{3}a_0\gamma_i/2$, where $a_0 = 2.46 $ \AA~  is the lattice constant of monolayer graphene, and the hoping amplitudes $\gamma_{i = 0,1,2,3,4}$ are defined as in Ref.~\cite{Zhang:2010aa}. $\delta_1$ denotes the average electrostatic potential difference between the middle and outer graphene layers, which arises from the difference of intrinsic chemical environment between the middle and outer graphene layers. $\delta_2$ denotes the on-site energy difference between the
low-energy sites ($A_1$ and $B_3$) and the high-energy sites ($B_1$, $A_2$,  $B_2$, and $A_3$) \cite{Zhang:2010aa}. Here we use $\Delta_d$ (instead of $U$ as in Ref.~\cite{Huang:2022aa}) to denote the potential difference between adjacent graphene layers induced by an 
externally applied displacement field. 
In the present study, we choose the same model parameters as given in Ref.~\cite{Zhou:2021aa,Zhou:2021ab} and reproduce them in Table.~\ref{tab:table1}.

\begin{table}
\centering
\caption{Summary of the model parameters used in this work \cite{Zhou:2021aa}. }
\label{tab:table1} 
\begin{tabular}{ c c c c c | c c }
\hline
\hline
 $\gamma_0$ (eV)&  $\gamma_1$ (eV) &  $\gamma_2$ (eV) &  $\gamma_3$ (eV)& $ \gamma_4$ (eV) & $\delta_1$ (meV) & $\delta_2$ (meV)  \\
 \hline
 3.1 & 0.38 & -0.015 & -0.29 & -0.141 & -2.3 & -10.5 \\
 \hline
 \hline
\end{tabular}
\end{table}

\section{Model Parameters near the Superconducting Dome}
\begin{figure}
 \centering
\includegraphics[width=0.75\columnwidth]{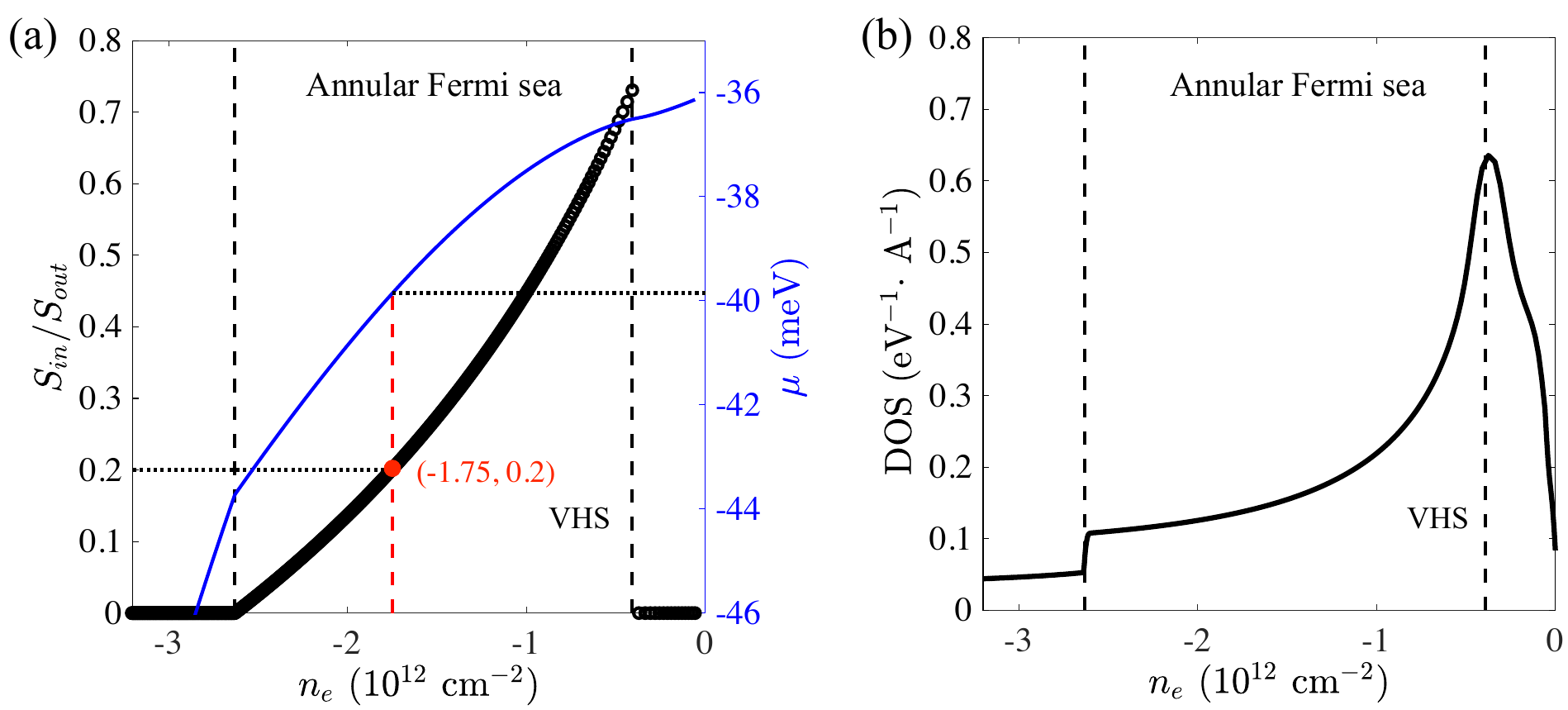}
 \caption{(a) $S_{in}/S_{out}$ and chemical potential $\mu$ versus carrier density $n_e$ at 
 $\Delta_d = 30$ meV.  The red dot marks the experimental value of $n_e$. 
 At this point, $\mu \sim -3.5$ meV measured from the VHS.  The value of 
 $\Delta_d$ has been adjusted so that $S_{in}/S_{out}$ is also consistent with experiment.
 (b) Density of states (DOS) versus $n_e$
 for $\Delta_d = 30$ meV.}
 \label{fig:figureS1}
\end{figure}

In RTG, the stronger superconducting (SC) dome (marked as SC1 in Ref.~\cite{Zhou:2021aa}) occurs close to a line in displacement field $D$ and carrier density $n_e$ experimentally controled parameter space. 
The SC1 phase emerges from a flavor-symmetric paramagnetic annular Fermi sea as revealed by normal-state quantum oscillation measurements. In this study, the single-particle energy spectrum of RTG is given by Eq.~(\ref{eq:bandmodel}), where $\Delta_d$ accounts for the screened electrostatic potential induced by $D$. 
To accurately capture the normal-state properties of the SC1 phase, we tune $\Delta_d$ and $\mu$ in Eq.~(\ref{eq:bandmodel}) until both of the quantum oscillation frequency $f_v$ and carrier density $n_e$ are consistent with experimentally reported data. Specifically, we have to solve the following equations
\begin{equation}
\begin{aligned}
f_v(\Delta_d,\mu) &= \frac{S_{v}}{|n_e|} = \frac{S_{v}}{4(S_{out}-S_{in})}, \\
n_e (\Delta_d,\mu) &= -1.75 \times 10^{12} \text{cm}^{-2},
\end{aligned}
\end{equation}
where $S_{v}$ denote the area of the momentum space enclosed by Fermi surface $v$, 
the areas enclosed by the larger and smaller Fermi surfaces of the annular Fermi sea are denoted by $S_{out}$ and $S_{in}$, respectively. For example, superconductivity was observed at $n_e \sim -1.75 \times 10^{12}$ cm$^{-2}$,
and $D=0.4$ V/nm.
Experiment shows that for these control parameters
$f_{out} \sim 0.31$, suggesting $S_{in}/S_{out} \sim 0.2$. As shown in Fig.~\ref{fig:figureS1}(a), we find both 
$n_e $ and $S_{in}/S_{out} $ are consistent with experimental data if we choose $\Delta_d = 30$ meV and $\mu = -3.5$ meV relative to the saddle-point van Hove singularity (VHS). 
The corresponding annular Fermi surfaces are given in Fig.~1(a) of the main text.
We choose these model parameters as a typical example for carrying out functional renormalization group (FRG) studies. 

We note that the bare electric potential generated by the displacement field $D\times d = 140~\text{meV}$, where $d = 0.35$ nm denotes inter-layer distance, is much larger than the value we used to fit the experimental data of $n_e$ and $f_{out}$. This is due to the screening effect, which has been systematically discussed in the supplementary materials of Ref.\cite{Huang:2022aa}.

\section{FRG flow equation}
As introduced in the main text, we employ the $N$-patch FRG method to investigate the low-temperature instabilities of the annular Fermi sea in RTG. The patch scheme is specified in Fig.~1(a), in which the momentum space around each valley is divided into $N = 48$ patches, implying that a total of $(2N)^3 = 884736$ 4-point vertex (4PV) functions 
must be calculated in each step of the FRG. The diagram representation of the one-loop FRG temperature-flow equations for the 4PVs is illustrated in Fig.~1(b) of the main text.  The corresponding algebraic equations are \cite{Honerkamp:2001aa}
\begin{equation}
\begin{aligned}
\partial_{T}u(\tau_1n_1,\tau_2n_2;\tau_3n_3)&=-\sum_{ \tau n} \partial_{T} \Pi^{pp}(\tau n, \bm{q}_{pp}) u(\tau_2n_2,\tau_1n_1;\tau n)u^*(\tau_3 n_3,\tau_4 n_4;\tau n)\\
&~~~-\sum_{\tau n} \partial_{T} \Pi^{ph}(\tau n, \bm{q}_{ex}) u^*(\tau_3 n_3,\tau n;\tau_1 n_1)u(\tau_2 n_2,\tau n;\tau_4 n_4) \\
&~~~+\sum_{\tau n }\partial_{T} \Pi^{ph}(\tau n, \bm{q}_{fs}) [2u^*(\tau n,\tau_4 n_4;\tau_1 n_1)u(\tau n,\tau_2 n_2;\tau_3 n_3)\\
&~~~~~~~~~~~~-u^*(\tau_4 n_4,\tau n;\tau_1 n_1)u(\tau n,\tau_2 n_2;\tau_3 n_3)-u^*(\tau n,\tau_4 n_4;\tau_1 n_1)u(\tau_2 n_2,\tau n;\tau_3 n_3)],
\end{aligned}
\label{eq:FRGflow}
\end{equation}
where $T$ is temperature, $u(\tau_1n_1,\tau_2n_2;\tau_3n_3)$ denotes a general 4PV defined on the Fermi surfaces, $\tau_{i} = \pm 1$ with $i = 1,\cdots,4$ distinguishes $K$ and $K'$ valley, and $n_{i}$ is the patch index (see Fig.~1(a) in the main text). The fourth patch and valley indices of $u(\tau_1n_1,\tau_2n_2;\tau_3n_3)$ is
determined by momentum and valley conservation:
\begin{equation}
\begin{aligned}
\bm{k}_{4} &= \bm{k}_{\tau_1 n_1}+\bm{k}_{\tau_2  n_2}-\bm{k}_{\tau_3 n_3}, \\
\tau_4 &= \tau_1+\tau_2-\tau_3.
\end{aligned}
\end{equation}
Here $\bm{k}_{\tau_i n_i}$ denotes the wavevector of patch $n_i$ on $\tau_i$-valley Fermi surfaces with respect to the center of  $\tau_i$ valley. Usually, $\bm{k}_4$ is not on the $\tau_4$-valley Fermi surface, we approximate $\bm{k}_4 $ by $ \bm{k}_{\tau_4 n_4}$ using the patch scheme developed in the main text. 
In Eq.~(\ref{eq:FRGflow}), the wavevector transfer for particle-particle scattering ($\bm{q}_{pp}$), exchange scattering ($\bm{q}_{ex}$), and forward scattering ($\bm{q}_{fs}$) are defined as
\begin{equation}
\begin{aligned}
\bm{q}_{pp}&=(\bm{k}_{\tau_1n_1}+\bm{k}_{\tau_2 n_2}) +(\tau_1+\tau_2)\bm{K}, \\
\bm{q}_{ex}&=(\bm{k}_{\tau_1 n_1}-\bm{k}_{\tau_3 n_3})+ (\tau_1-\tau_3)\bm{K}, \\
\bm{q}_{fs} &=(\bm{k}_{\tau_3 n_3}-\bm{k}_{\tau_2 n_2}) + (\tau_3-\tau_2)\bm{K}, 
\end{aligned}
\label{eq:transferq}
\end{equation}
where $\bm{K}$ is the wavevector of K point with respect to the BZ center. In Eq.~(\ref{eq:FRGflow}), $\Pi^{pp}(\tau n, \bm{q})$ and $\Pi^{ph}(\tau n, \bm{q})$ are non-interacting static particle-particle and particle-hole susceptibilities, defined as
\begin{equation}
\begin{aligned}
\Pi^{pp}(\tau n, \bm{q})&= \frac{1}{A} \sum_{\bm{k} \in \{ \tau , n \}}\frac{1-f(\epsilon_{\tau, \bm{k}})-f(\epsilon_{\tau',\bm{q}-\bm{k}})}{\epsilon_{\tau,\bm{k}}+\epsilon_{\tau',\bm{q}-\bm{k}}}, \\
\Pi^{ph}(\tau n, \bm{q})&= \frac{1}{A} \sum_{\bm{k} \in \{ \tau , n \}}\frac{f(\epsilon_{\tau,\bm{k}})-f(\epsilon_{\tau',\bm{k}-\bm{q}})}{\epsilon_{\tau,\bm{k}}-\epsilon_{\tau',\bm{k}-\bm{q}}}.
\end{aligned}
\label{eq:PPPHS1}
\end{equation}
where $A$ denotes the area of the system, $f(\epsilon)$ is the Fermi-Dirac distribution function, and $\epsilon_{\tau,\bm{k}}$ is the electron energy at valley $\tau$ and wavevector $\bm{k}$. In the above equation, summation of $\bm{k}$ is over patch regime $n$ in valley $\tau$, and valley $\tau'$ is determined by wavevector transfer $\bm{q}$  which may cross two valleys. 

The FRG flow equations can be simplified by employing time-reversal $\mathcal{T}$ symmetry, which requires 
\begin{equation}
\begin{aligned}
u(n_1\tau_1,n_2\tau_2;n_3\tau_3) &=  u^*(\bar{n}_1\bar{\tau}_1,\bar{n}_2\bar{\tau}_2;\bar{n}_3\bar{\tau}_3) \\
\Pi^{pp(h)}(n\tau,\bm{q}) &= \Pi^{pp(h)}(\bar{n}\bar{\tau},-\bm{q}),
\end{aligned}
\label{eq:time-reversal}
\end{equation}
where $\bar{n}$ denotes the opposite patch of $n$ on opposite valley of $\tau$ denoted by $\bar{\tau}$, indicating $\bm{k}_{\bar{\tau} \bar{ n}}= -\bm{k}_{\tau n} $. In the main text, the 4PVs are classified into intra-valley ($u_a$), inter-valley ($u_t$), and valley-exchange ($u_e$) types. Here we reproduce the definitions as follows
\begin{equation}
\begin{aligned}
u_a(n_1,n_2;n_3) &= u(K n_1,Kn_2;Kn_3) = u^*(K' \bar{n}_1,K'\bar{n}_2;K'\bar{n}_3), \\ 
u_t(n_1,n_2;n_3) &= u(Kn_1,K'n_2;K'n_3) = u^*(K'\bar{n}_1,K\bar{n}_2;K\bar{n}_3), \\
u_e(n_1,n_2;n_3) &= u(Kn_1,K'n_2;Kn_3) = u^*(K'\bar{n}_1,K\bar{n}_2;K'\bar{n}_3),
\end{aligned}
\label{eq:classify}
\end{equation}
where $\mathcal{T}$ symmetry is employed, and patch indices in $u_{a,t,e}$ are now associated with given valleys. By substituting Eqs.~(\ref{eq:transferq})-(\ref{eq:classify}) into Eq.~(\ref{eq:FRGflow}), we have the following set of RG flow equations
\begin{equation}
\begin{aligned}
\partial_{T}u_a(n_1,n_2;n_3) &=-\sum_{n} \partial_{T} \Pi^{pp}_{KK'}(n; \bm{k}_{n_1}+\bm{k}_{n_2}) u_a(n_2,n_1;n)u_a^*(n_3,n_4;n)\\
&~~~-\sum_{n} \partial_{T} \Pi^{ph}_{KK}(n; \bm{k}_{n_1}-\bm{k}_{n_3}) u_a^*(n_3,n;n_1)u_a(n_2,n;n_4) \\
&~~~-\sum_{n} \partial_{T} \Pi^{ph}_{KK}(n; \bm{k}_{n_3}-\bm{k}_{n_1}) u_e^*(n_3,\bar{n};n_1)u_e(n_2,\bar{n};n_4) \\
&~~~+\sum_{n}\partial_{T} \Pi^{ph}_{KK}(n; \bm{k}_{n_3}-\bm{k}_{n_2}) [2u_a^*(n,n_4;n_1)u_a(n,n_2;n_3)\\
&~~~~~~~~~~~~-u_a^*(n_4,n;n_1)u_a(n,n_2;n_3)-u_a^*(n,n_4;n_1)u_a(n_2,n;n_3)]\\
&~~~+\sum_{n}\partial_{T} \Pi^{ph}_{KK}(n;\bm{k}_{n_2}-\bm{k}_{n_3}) [2u_t(n,\bar{n}_4;\bar{n}_1)u_t^*(n,\bar{n}_2;\bar{n}_3)\\
&~~~~~~~~~~~~-u_e^*(n_4,\bar{n};n_1)u_t^*(n,\bar{n}_2;\bar{n}_3)-u_t(n,\bar{n}_4;\bar{n}_1)u_e(n_2,\bar{n};n_3)],
\end{aligned}
\label{eq:uaflow}
\end{equation}
\begin{equation}
\begin{aligned}
\partial_{T}u_t(n_1,n_2;n_3)&=-\sum_{n} \partial_{T} \Pi^{pp}_{KK}(n; \bm{k}_{n_1}-\bm{k}_{\bar{n}_2})u_t^*(\bar{n}_2,\bar{n}_1;\bar{n})u_t(\bar{n}_3,\bar{n}_4;\bar{n})\\
&~~~-\sum_{n} \partial_{T} \Pi^{pp}_{KK}(n;\bm{k}_{\bar{n}_2}-\bm{k}_{n_1})u_e^*(\bar{n}_2,\bar{n}_1;n)u_e(\bar{n}_3,\bar{n}_4;n)\\
&~~~-\sum_{n} \partial_{T} \Pi^{ph}_{KK'}(n; \bm{k}_{n_1}+\bm{k}_{\bar{n}_3}) u_t(\bar{n}_3,\bar{n};\bar{n}_1)u_t^*(\bar{n}_2,\bar{n};\bar{n}_4) \\
&~~~+\sum_{n}\partial_{T} \Pi^{ph}_{KK}(n; \bm{k}_{\bar{n}_2}-\bm{k}_{\bar{n}_3}) [2u_a^*(n,n_4;n_1)u_t(n,n_2;n_3)\\
&~~~~~~~~~~~~-u_a^*(n_4,n;n_1)u_t(n,n_2;n_3)-u_a^*(n,n_4;n_1)u_e^*(\bar{n}_2,\bar{n};\bar{n}_3)]\\
&~~~+\sum_{n}\partial_{T} \Pi^{ph}_{KK'}(n;\bm{k}_{\bar{n}_3}-\bm{k}_{\bar{n}_2}) [2u_t(n,\bar{n}_4;\bar{n}_1)u_a^*(n,\bar{n}_2;\bar{n}_3)\\
&~~~~~~~~~~~~-u_e^*(n_4,\bar{n};n_1)u_a^*(n,\bar{n}_2;\bar{n}_3)-u_t(n,\bar{n}_4;\bar{n}_1)u_a^*(\bar{n}_2,n;\bar{n}_3)],
\end{aligned}
\label{eq:utflow}
\end{equation}
\begin{equation}
\begin{aligned}
\partial_{T}u_e(n_1,n_2;n_3) &=-\sum_{n} \partial_{T} \Pi^{pp}_{KK}(n; \bm{k}_{n_1}-\bm{k}_{\bar{n}_2}) u_t^*(\bar{n}_2,\bar{n}_1;\bar{n})u_e^*(n_3,n_4;n)\\
&~~~-\sum_{n} \partial_{T} \Pi^{pp}_{KK}(n; \bm{k}_{\bar{n}_2}-\bm{k}_{n_1}) u_e^*(\bar{n}_2,\bar{n}_1;n)u_t^*(n_3,n_4;\bar{n})\\
&~~~-\sum_{n} \partial_{T} \Pi^{ph}_{KK}(n; \bm{k}_{n_1}-\bm{k}_{n_3}) u_a^*(n_3,n;n_1)u_e^*(\bar{n}_2,\bar{n};\bar{n}_4) \\
&~~~-\sum_{n} \partial_{T} \Pi^{ph}_{KK}(n; \bm{k}_{n_3}-\bm{k}_{n_1}) u_e^*(n_3,\bar{n};n_1)u_a^*(\bar{n}_2,n;\bar{n}_4) \\
&~~~+\sum_{n}\partial_{T} \Pi^{ph}_{KK'}(n; \bm{k}_{n_3} + \bm{k}_{\bar{n}_2}) [2u_e^*(n,n_4;n_1)u_e(n,n_2;n_3)\\
&~~~~~~~~~~~~-u_t(\bar{n}_4,\bar{n};\bar{n}_1)u_e(n,n_2;n_3)-u_e^*(n,n_4;n_1)u_t^*(\bar{n}_2,\bar{n};\bar{n}_3)].
\end{aligned}
\label{eq:ueflow}
\end{equation}
In the above equations, the patch wavevectors $\bm{k}_{n_i}$ and $\bm{k}_{\bar{n}_i}$ are now defined on the $K$-valley Fermi surfaces with respect to $K$-valley center. The intra-and inter-valley particle-particle and particle-hole susceptibilities are given as 
\begin{equation}
\begin{aligned}
\Pi^{pp}_{KK}(n; \bm{q}) & = \frac{1}{A}\sum_{\bm{k}\in \{K,n\}} \frac{1-f(\epsilon_{\bm{k}})-f(\epsilon_{\bm{k}-\bm{q}})}{\epsilon_{\bm{k}} + \epsilon_{\bm{k}-\bm{q}}}, \\
\Pi^{pp}_{KK'}(n; \bm{q}) & =\frac{1}{A}\sum_{\bm{k}\in \{K,n\}} \frac{1-f(\epsilon_{\bm{k}})-f(\epsilon_{\bm{q}-\bm{k}})}{\epsilon_{\bm{k}} + \epsilon_{\bm{q}-\bm{k}}}, \\
\Pi^{ph}_{KK}(n;\bm{q}) & =  \frac{1}{A}\sum_{\bm{k}\in \{K,n\}} \frac{f(\epsilon_{\bm{k}})-f(\epsilon_{\bm{k}-\bm{q}})}{\epsilon_{\bm{k}} - \epsilon_{\bm{k}-\bm{q}}} , \\
\Pi^{ph}_{KK'}(n;\bm{q}) & =\frac{1}{A}\sum_{\bm{k}\in \{K,n\}} \frac{f(\epsilon_{\bm{k}})-f(\epsilon_{\bm{q}-\bm{k}})}{\epsilon_{\bm{k}} - \epsilon_{\bm{q}-\bm{k}}}, \\ 
\end{aligned}
\end{equation}
where $\epsilon_{\bm{k}}$ is the simplified notation for $K$-valley energy spectrum $\epsilon_{K,\bm{k}}$, and we have used the relation $\epsilon_{K,\bm{k}} = \epsilon_{K',-\bm{k}}$ due to $\mathcal{T}$ symmetry. We note that the fourth patch index in $u_a(n_1,n_2;n_3)$, $u_t(n_1,n_2;n_3)$ and $u_e(n_1,n_2;n_3)$ shown in Eqs.~(\ref{eq:uaflow})-(\ref{eq:ueflow}) are obtained respectively by
\begin{equation}
\begin{aligned}
u_a:&~~~~\bm{k}_{n_1}+\bm{k}_{n_2}-\bm{k}_{n_3} \rightarrow \bm{k}_{n_4}, \\
u_t:&~~~~\bm{k}_{n_1}-\bm{k}_{\bar{n}_2}+\bm{k}_{\bar{n}_3} \rightarrow \bm{k}_{n_4}, \\
u_e:&~~~~-\bm{k}_{n_1}+\bm{k}_{\bar{n}_2}+\bm{k}_{n_3} \rightarrow \bm{k}_{\bar{n}_4},
\end{aligned}
\end{equation}
where $\rightarrow$ approximates the vectors on the left hand side to the patch wavevectors on the right hand side using the patch scheme developed in the main text. By taking advantage of $\mathcal{T}$ symmetry, we demonstrate that the valley-index can be formally removed from the RG flow equations as given in Eqs.~(\ref{eq:uaflow})-(\ref{eq:ueflow}). Therefore, we only need to calculate $3N^3 =  331776$ 4PV functions in each step of FRG. 

In the temperature-flow RG scheme \cite{Honerkamp:2001aa}, the momentum-space structure of the 4PV develops quickly at low temperatures because the relevant particle-particle and particle-hole susceptibilities become more 
sensitive to the details of electronic structure.  To enhance the efficiency of numerical calculation, we choose dimensionless parameter $y = \ln{(\Lambda/k_BT)}$ as the RG flow time, where $\Lambda = 1$ eV is the ultraviolet energy cut-off. Here $y=0$ corresponds to $T \sim 1.16 \times 10^4$ K, above which contributions from both of particle-particle and particle-hole fluctuations to RG flows are negligible. 
In the main text, we have transformed the RG flow time to temperature.

\section{Coulomb interaction}
\label{sec:Coulomb}
In this study, we consider the long-range Coulomb interactions
\begin{equation}
H_{ee} = \frac{1}{2} \sum_{ll'\tau\tau'}\sum_{\bm{q}} v_{ll'}(\bm{q}) \hat{\rho}_{l \tau \tau}(\bm{q}) \hat{\rho}_{l' \tau' \tau'}(-\bm{q}) 
+ \frac{1}{2} \sum_{ll'\tau}\sum_{\bm{q}} v_{ll'}(\bm{Q}) \hat{\rho}_{l \tau \bar{\tau}}(\bm{q}) \hat{\rho}_{l' \bar{\tau} \tau}(-\bm{q}).
\label{eq:Coulomb}
\end{equation}
The second term on the right hand of the above equation accounts for
valley-exchange interactions, $\bar{\tau}$ denotes the opposite valley to $\tau$, 
$\bm{Q}$ is the momentum-difference between two valleys, 
and the valley-dependent density matrix is defined as 
\begin{equation}
\hat{\rho}_{l\tau\tau'}(\bm{q}) = \sum_{\sigma \bm{k}} \psi_{l \tau \sigma \bm{k}}^{\dagger}\psi_{ l \tau' \sigma \bm{k}+\bm{q}}.
\end{equation}
In typical dual-gated devices, the screened Coulomb potential 
\begin{equation}
v_{ll'}(\bm{q}) = \frac{2\pi e^2}{\epsilon q}e^{-q d|l-l'|}\tanh{(\sqrt{\frac{\epsilon_{\|}}{\epsilon_{\bot}}}qd_s)}
\end{equation}
where the interlayer spacing $d \sim 0.35 $ nm and the separation between the metallic gates and 
the RTG sample $d_s$ has typical values $\sim 30 - 60$ nm. 
For the commonly employed hexagonal boron nitride (hBN) dielectric insulator, 
the in-plane and vertical components of its dielectric tensor 
are $\epsilon_{\|} \sim 6.9 $ and $\epsilon_{\bot} \sim 3.48$ \cite{Laturia:2018aa}, giving $\epsilon = \sqrt{\epsilon_{\|} \epsilon_{\bot} } \sim 5$. In Eq.~(\ref{eq:Coulomb}), the valley-exchange interaction is much weaker than intra- and inter-valley 
interaction (see definitions in Fig.~2(a) of the main text) because the magnitude of $|\bm{Q}| = 2|\bm{K}| \sim 34.05$ nm$^{-1}$ is much larger than the typical wavevector transfer $|\bm{q}| \lesssim 0.7$ nm$^{-1}$ within the annular Fermi surfaces. 

\begin{figure}
 \centering
\includegraphics[width=\columnwidth]{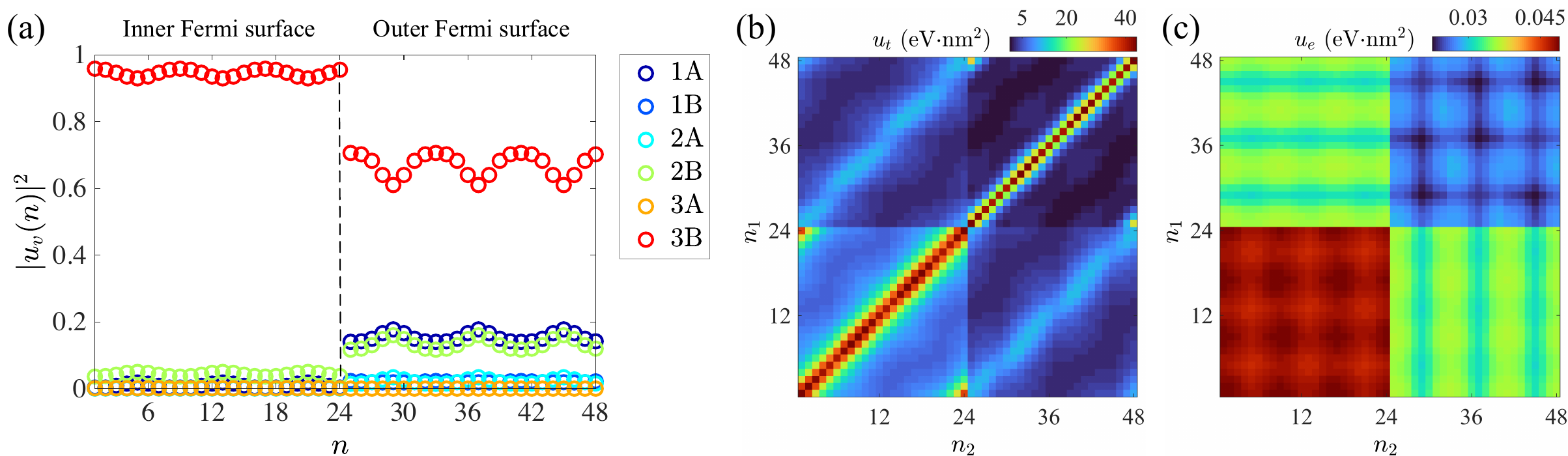}
 \caption{(a) Layer and sublattice components of valence band wave functions at patch wavevectors. (b)-(c) Initial values of pairing interactions (b) $u_t(n_1,\bar{n}_1;\bar{n}_2)$ and (c) $u_e(n_1,\bar{n}_1;n_2)$. Similar to Fig.~2(b)-(c) of the main text, the bottom-left and top-right 24$\times$24 blocks are scatterings within the inner and outer Fermi surfaces, respectively. The remaining blocks are scatterings across the annular Fermi surfaces. }
 \label{fig:figureS2}
\end{figure}

The initial values of the 4PVs are obtained by projecting the bare Coulomb interaction onto the valence band,
\begin{equation}
\begin{aligned}
u_a(n_1,n_2;n_3)  &  = \sum_{ll'} v_{ll'}(\bm{k}_{n_2}-\bm{k}_{n_3}) \langle u_{v}(\bm{k}_{n_1})|u_{v}(\bm{k}_{n_4})\rangle_{l}\langle u_{v}(\bm{k}_{n_2})|u_{v}(\bm{k}_{n_3})\rangle_{l'}, \\
u_t(n_1,n_2;n_3) & = \sum_{ll'} v_{ll'}(\bm{k}_{\bar{n}_2}-\bm{k}_{\bar{n}_3}) \langle u_{v}(\bm{k}_{n_1})|u_{v}(\bm{k}_{n_4})\rangle_{l} \langle u_{v}(\bm{k}_{\bar{n}_3})|u_{v}(\bm{k}_{\bar{n}_2})\rangle_{l'}, \\
u_e(n_1,n_2;n_3)  & =  \sum_{ll'} v_{ll'}(\bm{Q})    \langle u_{v}(\bm{k}_{n_1})|u^*_{v}(\bm{k}_{\bar{n}_4})\rangle_{l} \langle u^*_{v}(\bm{k}_{\bar{n}_2})|u_{v}(\bm{k}_{n_3})\rangle_{l'},
\end{aligned}
\label{eq:Coulombmatrix}
\end{equation}
where $\langle u_{v}(\bm{k}_{n_1})|u_{v}(\bm{k}_{n_4})\rangle_{l}$ is the layer-resolved overlap between valence band Bloch wave functions $|u_{v}(\bm{k}_{n_1}) \rangle$ and $| u_{v}(\bm{k}_{n_4}) \rangle$.
Figure.~\ref{fig:figureS2}(a) plots layer and sublattice-resolved components of $|u_{v}(\bm{k}_{n}) \rangle$ at patch wavecector $\bm{k}_{n}$. The wave functions on the inner Fermi surface are largely localized on B sites of the third graphene layer. Due to the increased inter-layer coupling at larger wavevector $\bm{k}$, the wave functions on the outer (larger) Fermi surface are more extended into other graphene layers. For a given graphene layer, the wave functions on both of the inner and outer Fermi surfaces are nearly sublattice polarized.

To eliminate the artificial divergence of the intra-patch Coulomb potential 
introduced by patching momentum space, we approximate the Coulomb potential in Eq.~(\ref{eq:Coulombmatrix}) using patch-averaged wavevector transfers. Taking $u_a(n_1,n_2;n_3,n_4)$ as an example, we first generate a large 
number of random $\bm{k}$ points within each patch regime $n_{i=1,2,3,4}$, and perform the following averaging
\begin{equation}
\frac{1}{\bar{q}_{23} }= \sum_{i\in n_2}\sum_{j\in n_3} \frac{e^{-d |\bm{k}_i-\bm{k}_j| |l-l'|}}{|\bm{k}_i-\bm{k}_j|} \tanh{(\sqrt{\frac{\epsilon_{\|}}{\epsilon_{\bot}}} |\bm{k}_i-\bm{k}_j|d_s)},
\end{equation}
where $\bm{q}_{i,j}$ are random wavevectors generated within patch $n_{2,3}$, and $\bar{q}_{23}$ depends on layer indices $l$ and $l'$. Similar averaging can be done for $\bar{q}_{14}$. We then approximate the relevant Coulomb potential in $u_a(n_1,n_2;n_3,n_4)$ by
\begin{equation}
 v_{ll'}(\bar{q}) = \frac{2\pi e^2}{\epsilon \bar{q}},
\end{equation}
where the averaged wavevector transfer $\bar{q} =(\bar{q}_{23}  + \bar{q}_{14} )/2 $. Finally, by including the wave function effects, we have 
\begin{equation}
u_a(n_1,n_2;n_3)    = \sum_{ll'} v_{ll'}(\bar{q}) \langle u_{v}(\bm{k}_{n_1})|u_{v}(\bm{k}_{n_4})\rangle_{l}\langle u_{v}(\bm{k}_{n_2})|u_{v}(\bm{k}_{n_3})\rangle_{l'}.
\end{equation}
Similar procedures can be applied to $u_t(n_1,n_2;n_3,n_4)$. Figure.~\ref{fig:figureS2}(b)
 and (c) plot the initial values of $u_t$ and $u_e$ in the opposite-valley pairing channel, where $u_e$ is $\sim$100 times weaker than that of inter-valley interaction $u_t$. In Fig.~2(b)-(c) of the main text, we show that $u_e$ strengthens under RG and flows to comparable values to $u_t$. 
 
Based on the above analyses, the initial value of a 4PV is in general a complex number because the band eigenstates $|u_{v}(\bm{k}_{n_1}) \rangle$ are six-component layer/sublattice vectors. Nevertheless, due to the presence of time-reversal symmetry, the pairing interaction $u_t(n_1,\bar{n}_1;\bar{n}_2)$ and $u_e(n_1,\bar{n}_1;n_2)$ (see Figs.~\ref{fig:figureS2}(b)-(c) and Figs.~2(b)-(c) in the main text) are real because scatterings occur between two time-reversal partners.
 
The FRG calculations in the this work do not include the electron-phonon interactions for two reasons. First, the acoustic and optical phonon-mediated intra- and inter-valley interaction \cite{Wu:2018aa,Chou:2021aa} is weak compared to the valence band projected bare Coulomb interaction given in Fig.~\ref{fig:figureS2}(b). Secondly, the in-plane optical phonon-mediated valley-exchange interaction involves sublattice flipping \cite{Wu:2018aa}. Therefore it will be dramatically suppressed in RTG because the wave functions on the annular Fermi surfaces are nearly sublattice polarized, as shown in Fig.~\ref{fig:figureS2}(c).

\section{Chiral superconducting states}
\label{sec:chiral}
In the main text, we show that the renormalized pairing interactions result in doubly-degenerate $p$-wave-like pairing in spin-triplet channel and doubly-degenerate $d$-wave-like pairing in spin-singlet channel. Below the critical temperature $T_c$, the degeneracy may be lifted to a particular linear combination of the doubly degenerate eigenstates. To identify which kind of linear combination is energetically more favorable, we may calculate the corresponding superconducting condensation energy by solving the nonlinear gap equation. Here we employ an alternative approach to extract the superconducting free energy by integrating out the fermionic fields, and then expand the free energy up to quartic orders.

\subsection{Mean-field Hamiltonian}
Based on the results of FRG calculations, the pairing interaction Hamiltonian
\begin{equation}
\begin{aligned}
H_{int} &=  \sum_{n_1 n_2  ss'}  u_t(n_1,\bar{n}_1;\bar{n}_2,n_2)
\psi^{\dagger}_{Kn_1s} \psi^{\dagger}_{K'\bar{n}_1 s'}   \psi_{K'\bar{n}_2 s'} \psi_{Kn_2 s} + u_e(n_1,\bar{n}_1;n_2,\bar{n}_2) \psi^{\dagger}_{K n_1 s} \psi^{\dagger}_{K' \bar{n}_1 s'}  \psi_{K n_2 s'}  \psi_{K' \bar{n}_2 s}, 
\end{aligned}
\end{equation}
where $n_{1,2}$ are patch indices, $s,s'$ denote electron spin. 
Within mean-field theory, the pairing potential
\begin{equation}
\Delta_{ss'}(n_1) = \sum_{n_2}  u_t(n_1,\bar{n}_1;\bar{n}_2,n_2)F_{ss'}(n_2) - u_e(n_1,\bar{n}_1;n_2,\bar{n}_2) F_{s's}(n_2) 
\end{equation}
where $F_{ss'} (n_2)=  \langle  \psi_{K'\bar{n}_2 s'} \psi_{Kn_2 s}  \rangle $. The spin-singlet pairing potential is constructed as
\begin{equation}
 \Delta_{s}(n_1) =  \frac{1}{2}[\Delta_{\uparrow \downarrow} (n_1)-\Delta_{\downarrow\uparrow }(n_1) ]= \sum_{n_2}  V_s(n_1;n_2)[ F_{\uparrow \downarrow}(n_2)-F_{\downarrow \uparrow}(n_2)]/2,
\end{equation}
and spin-triplet pairing potential 
\begin{equation}
\begin{aligned}
 \Delta_{t}(n_1) &= \frac{1}{2}[\Delta_{\uparrow \downarrow} (n_1)+\Delta_{\downarrow\uparrow }(n_1) ] = \sum_{n_2}  V_t(n_1;n_2) [ F_{\uparrow \downarrow}(n_2)+F_{\downarrow \uparrow}(n_2)]/2  \\
  \Delta_{t}(n_1) & = \Delta_{\uparrow \uparrow, \downarrow \downarrow } (n_1)= \sum_{n_2}  V_t(n_1;n_2)  F_{\uparrow \uparrow, \downarrow \downarrow}(n_2).
\end{aligned}
\end{equation}
The pairing interaction in spin-singlet and spin-triplet channels are defined as
\begin{equation}
V_{s,t}(n_1;n_2) = u_t(n_1,\bar{n}_1;\bar{n}_2,n_2)\pm  u_e(n_1,\bar{n}_1;n_2,\bar{n}_2).
\end{equation}
In the main text, we show that the strongest pairing instabilities induced by $V_{s}$ and $V_{t}$ are doubly-generate $d$-wave-like and $p$-wave-like pairing, respectively. Here we choose spin-singlet $d$-wave pairing as an example to show which kind of linear combination between the two degenerate states is energetically more favorable below $T_c$. The mean-field Hamiltonian for spin-singlet pairing 
\begin{equation}
\begin{aligned}
H_{MF} &=  \sum_{n}  
 \Delta(n ) \psi^{\dagger}_{Kn \uparrow} \psi^{\dagger}_{K'\bar{n} \downarrow}  +\text{H.c.} -  \text{Tr} (\Delta^{\dagger} V^{-1}_s \Delta ).
\end{aligned}
\end{equation}
Here we omit subscript "$s$" in the spin-singlet paring potential, and use the definition $\Delta = \text{diag}[\Delta(n)]$. 

\subsection{Effective free energy}
\label{subsec:Effect}
Within the patch scheme developed in the main text, the action of the spin-singlet pairing state
\begin{equation}  
\begin{aligned}
S[\psi^{\dagger}, \psi] =& -\sum_{n } \int_{P} \psi^{\dagger}_{K n \uparrow,P} (i\omega-\epsilon_{ K n,\bm{k}}) \psi_{K n \uparrow,P} + \psi^{\dagger}_{K' \bar{n} \downarrow,-P} (-i\omega-\epsilon_{ K' \bar{n},-\bm{k}}) \psi_{K' \bar{n} \downarrow,-P} \\
 &+  \sum_{n}  \int_{P}
 \left[\psi^{\dagger}_{Kn\uparrow,P} \Delta(n) \psi^{\dagger}_{K'\bar{n} \downarrow,-P}  +\text{H.c.} \right] - \text{Tr} (\Delta^{\dagger} V^{-1}_s \Delta ) \\
 =&   - \int_{P} \Psi_{P}^{\dagger}  
 \begin{pmatrix}
 i\omega -\epsilon_{K,\bm{k}} & -\Delta \\
 -\Delta^{\dagger} &  i\omega + \epsilon_{K',-\bm{k}}
 \end{pmatrix}
 \Psi_{P}
 -\text{Tr} (\Delta^{\dagger} V^{-1}_s \Delta)
\end{aligned}
\end{equation}
where the Nambu spinor $\Psi_{P}^{\dagger}  = [\psi^{\dagger}_{K 1 \uparrow,P}, \psi^{\dagger}_{K 2 \uparrow,P}, \dots; \psi_{K' \bar{1} \downarrow,-P},\psi_{K' \bar{2} \downarrow,-P},\dots]$ and $\epsilon_{K,\bm{k}} = \text{diag}(\epsilon_{Kn,\bm{k}})$ denotes the patch-diagonal matrix of electron energy. The collective label $P = (i\omega,\bm{k})$ consists of fermionic Matsubara frequencies $i\omega$, and wavevector $\bm{k}$, and the integration over $P$ is defined as  
\begin{equation}
\int_{P} = \frac{k_BT}{ A}\sum_{i\omega} \sum_{\bm{k} \in \text{patch}~n}, 
\label{eq:intdefine}
\end{equation}
where $A$ is the area of the system. 
$\mathcal{T}$ symmetry requires $\epsilon_{K',-\bm{k}} = \epsilon_{K,\bm{k}}$. By integrating the fermionic fields, the partition function
\begin{equation}
\begin{aligned}
Z &= \int \mathcal{D}[\psi^{\dagger}, \psi] e^{-S[\psi^{\dagger}, \psi]} \\
& = e^{\text{Tr} (\Delta^{\dagger} V^{-1}_s \Delta )
+ \int_{P} \text{Tr} \ln
 \begin{pmatrix}
 i\omega +\epsilon_{K,\bm{k}} & \Delta \\
 \Delta^{\dagger} &  i\omega - \epsilon_{K,\bm{k}}
 \end{pmatrix}
},
\end{aligned}
\end{equation}
The Free energy is then obtained as
\begin{equation}
\mathcal{F} = - \text{Tr} [\Delta^{\dagger} V^{-1}_s \Delta ] - \int_{P} \text{Tr} \ln
 \begin{pmatrix}
 G^{-1}_{+} & \Delta \\
 \Delta^{\dagger} &  G^{-1}_{-}
 \end{pmatrix},
 \label{eq:freeenergy1}
\end{equation}
where the matrix form of Green's function $G_{+,-} = (i\omega \pm   \epsilon_{K,\bm{k} })^{-1}$. When temperature is slightly below $T_c$, perturbation expansion of Eq.~(\ref{eq:freeenergy1}) up to quartic terms of $\Delta$ reads
\begin{equation}
\begin{aligned}
\mathcal{F} &\approx - \text{Tr} (\Delta^{\dagger} V^{-1}_s \Delta ) +\frac{1}{2} \int_{P} \text{Tr} \begin{pmatrix}
 0 & \Delta G_{-}\\
 \Delta^{\dagger} G_{+} &  0
 \end{pmatrix}^2  
  +\frac{1}{4} \int_{P} \text{Tr} \begin{pmatrix}
 0 & \Delta G_{-}\\
 \Delta^{\dagger} G_{+} &  0
 \end{pmatrix}^4 \\
 & = - \text{Tr} (\Delta^{\dagger} V^{-1}_s \Delta )  +  \int_{P} \text{Tr}( \Delta G_{-} \Delta^{\dagger} G_{+}) + \frac{1}{2}
 \int_{P} \text{Tr}( \Delta G_{-} \Delta^{\dagger} G_{+} \Delta G_{-} \Delta^{\dagger} G_{+}) \\
 &=- \text{Tr} (\Delta^{\dagger} V^{-1}_s \Delta )   +   \text{Tr} \left( K_1   |\Delta|^2  \right)    +\frac{1}{2}\text{Tr} \left( K_2  |\Delta|^4 \right).  
 \end{aligned}
 \label{eq:freeenergyapp} 
\end{equation}
Here $K_{1,2}$ are patch-diagonal matrices defined as
\begin{equation}
\begin{aligned}
K_1 &= \int_{P} G_{-} G_{+} = k_BT\sum_{i\omega} \sum_{\bm{k} \in n} \frac{1}{ (i\omega +   \epsilon_{K,\bm{k} })(i\omega -   \epsilon_{K,\bm{k} })}=-\sum_{\bm{k} \in n} \frac{1-2f(\epsilon_{K,\bm{k}})}{2\epsilon_{K,\bm{k}}} =- \Pi^{pp}_{KK}(0) <0, \\
K_2 &= \int_{P} G_{-} G_{+} G_{-} G_{+} =  k_BT\sum_{i\omega} \sum_{\bm{k} \in n} \frac{1}{ (i\omega +   \epsilon_{K,\bm{k} })^2(i\omega -   \epsilon_{K,\bm{k} })^2} = \sum_{\bm{k} \in n} \frac{1-2f(\epsilon_{K,\bm{k}})}{4\epsilon^3_{K,\bm{k}}} -\frac{\beta}{2\epsilon^2_{K,\bm{k}}}f(\epsilon_{K,\bm{k}})f(-\epsilon_{K,\bm{k}})>0.
\end{aligned}
\label{eq:Kdefinition}
\end{equation}
In the above equations, $\Pi^{pp}_{KK}(0) = \text{diag}[\Pi^{pp}_{KK}(n,0)]$ are zero-$\bm{q}$ particle-particle susceptibilities.

\subsection{Pairing symmetry}
\label{subsec:chiral_pairing}
In the main text, we show that $V_s$ has two degenerate $d$-wave-like eigenvectors
\begin{equation}
V_{s} |d_{1,2} \rangle = V_s^d|d_{1,2} \rangle,
\label{eq:d-waveeignv}
\end{equation}
where $V_s^d$ is the corresponding eigenvalue, denoting effective pairing interaction in the spin-singlet $d$-wave channel. Due to $\mathcal{T}$ symmetry, $d_{1,2}$ can be chosen as real and orthogonal. For example,
\begin{equation}
V_{s} |d_{1,2}^* \rangle = V_s^d |d_{1,2}^* \rangle,
\end{equation}
and two normalized real eigenvectors can be constructed as
\begin{equation}
\begin{aligned}
|v_{1} \rangle = \frac{|d_{1} \rangle +|d_{1} ^*\rangle}{\sqrt{2+2\text{Re}(\langle d_1|d_1^*\rangle)}}, ~~
|v_{2} \rangle = \frac{|d_{2} \rangle +|d_{2} ^*\rangle}{\sqrt{2+2\text{Re}(\langle d_2|d_2^*\rangle)}}.
\end{aligned}
\end{equation}
Usually, $\langle v_{2}|v_{1} \rangle \neq 0$. Orthogonal eigenvectors are easily constructed by proper linear combinations of $v_{1}$ and $v_{2}$ using real coefficients. 

\begin{figure}
 \centering
\includegraphics[width=\columnwidth]{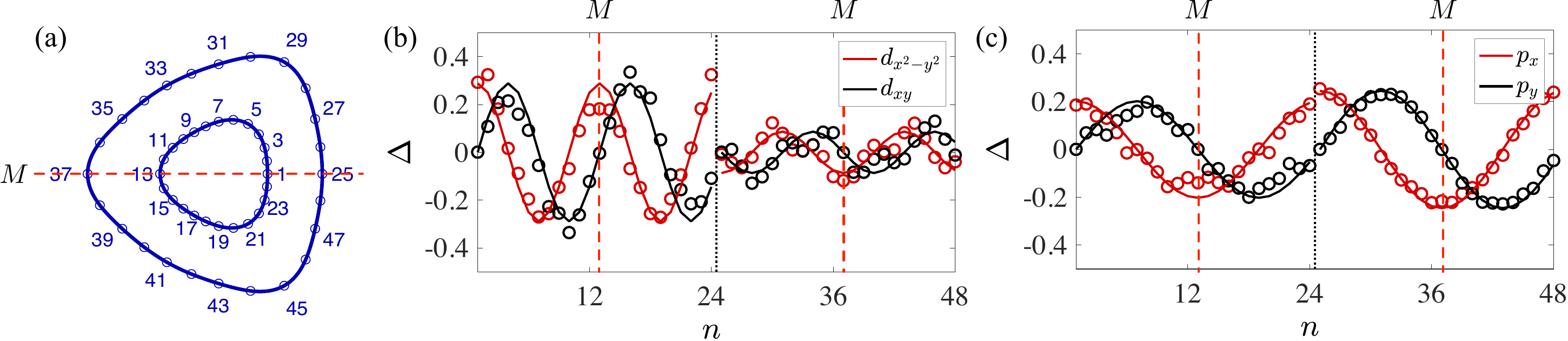}
 \caption{(a) $K$-valley Fermi surface, where red dashed line denotes one of the three vertical mirror planes. (b) Two degenerate $d$-wave-like eigenvectors (circles) of $V_s$. Solid curves are fittings to $d_{x^2-y^2}$ and $d_{xy}$-wave solutions for systems with full rotational symmetry. The $d_{x^2-y^2}$ and $d_{xy}$-wave-like solutions are respectively even and odd functions with respect to the mirror center marked by red dashed lines. (c) Two degenerate $p$-wave like eigenvectors (circles) of $V_t$. Solid curves are fittings to $p_{x}$ and $p_{y}$-wave solutions. The $p_{x}$ and $p_{y}$-wave-like solutions are respectively even and odd functions with respect to the mirror center marked by red dashed lines.}
 \label{fig:figureS3}
\end{figure}

In order to simplify the free energy in Eq.~(\ref{eq:freeenergyapp}), we construct the two-degenerate $d$-wave-like eigenvectors based on symmetry arguments. Since $V_s$ is invariant under mirror symmetry illustrated in Fig.~\ref{fig:figureS3}(a), the real eigenvector $|d_{1} \rangle$ can be chosen to be the eigenvector of mirror operator $M$ with eigenvalue $1$, namely, 
\begin{equation}
M |d_{1} \rangle =   |d_{1} \rangle
\label{eq:mirror_d1}
\end{equation}
One possible choosing of $|d_{1} \rangle$ is given in Fig.~\ref{fig:figureS3}(b), which is identified as $d_{x^2-y^2}$-like solution.
The patch representation of $M$ reads 
\begin{equation}
M = 
\begin{pmatrix}
I_{1\times 1} & 0 \\
0 & O_{23\times 23} 
\end{pmatrix} 
\end{equation}
where $I_{n\times n}$ denotes $n$-dimensional identity matrix, $O_{n\times n}$ denotes $n$-dimensional off-diagonal identity matrix, for example, $O_{2\times 2} = \begin{pmatrix}
0 &1\\
1& 0
\end{pmatrix}$. The degenerate eigenvector $|d_{2} \rangle$ can be constructed as
\begin{equation}
|d_{2} \rangle = \frac{1}{\sqrt{3}}(C_3-C_3^{-1})|d_{1} \rangle ,
\label{eq:d1d2relation}
\end{equation}
where $C_3$ denotes three-fold rotational operator, and
\begin{equation}
C_3 = 
\begin{pmatrix}
0 & I_{8\times 8} &0  \\
0 & 0 &I_{8\times 8} \\
I_{8\times 8} & 0 & 0  
\end{pmatrix}.
\end{equation}
From the definition, we have $C_3^2 = C_{3}^{-1} = C_{3}^{\text{T}}$. It is easy to verify the normalization and orthogonal conditions 
\begin{equation}
\begin{aligned}
\langle d_{2}|d_{2} \rangle &= \frac{1}{3} \langle d_{1}|(C_3^{-1}-C_3)(C_3-C_3^{-1}) |d_{1} \rangle = \frac{1}{3}\langle d_{1}|(2-C_3-C_3^2) |d_{1} \rangle = \langle d_{2}|d_{2} \rangle =1, \\
\langle d_{1}|d_{2} \rangle &= \frac{1}{\sqrt{3}} \langle d_{1}|(C_3-C_3^{-1}) |d_{1} \rangle =0,
\end{aligned}
\label{eq:nor_orth}
\end{equation}
where we have used the relation $(1+C_3+C_3^2) |d_{1} \rangle = 0 $. For a general matrix $U$ that is invariant under three-fold rotational symmetry, $C_3^{-1}UC_3 = U$, we have 
\begin{equation}
\langle d_{2} |U|d_{2} \rangle = \frac{1}{3} \langle d_{1}|(C_3^{-1}-C_3)U(C_3-C_3^{-1}) |d_{1} \rangle = \langle d_{1} |U|d_{1} \rangle
\label{eq:theorem1}
\end{equation}
therefore,
\begin{equation}
\langle d_{2}|V_{s} |d_{2} \rangle = \langle d_{1}|V_{s} |d_{1} \rangle = V_s^d 
\end{equation}
suggesting $|d_{1,2} \rangle$ is a real eigenvector of $V_s$ with eigenvalue $V_s^d$. Moreover, we have
\begin{equation}
M |d_{2} \rangle = \frac{1}{\sqrt{3}} M (C_3-C_3^{-1})|d_{1} \rangle = - |d_{2} \rangle.
\label{eq:mirror_d2}
\end{equation}
The constructed $|d_{2} \rangle$ is given in Fig.~\ref{fig:figureS3}(b), and identified as $d_{xy}$-like solution. 

\subsection{Chiral $d\pm id$ pairing}
A general $d$-wave-like pairing potential is expressed by the following linear combination
\begin{equation}
\Delta(n) = \Delta_{1} |d_{1}\rangle + \Delta_{2} |d_{2}\rangle,
\end{equation}
where $\Delta_{1,2}$ are two complex numbers, characterizing the components of $|d_{1}\rangle$ and $|d_{2}\rangle$. The normalization condition requires 
\begin{equation}
|\Delta_{1}|^2+|\Delta_{2}|^2 = N|\bar{\Delta}|^2,
\end{equation}
where $N = 48$ is the total number of patches on the valley-projected Fermi surface (see Fig.~1(a) in the main text), and $\bar{\Delta}$ denotes the averaged magnitude of the pairing potential over the annular Fermi surfaces. The matrix form of the pairing potential 
\begin{equation}
\Delta  =  \Delta_1 \mathcal{D}_1+\Delta_2 \mathcal{D}_2
\label{eq:linearDel}
\end{equation}
where  patch-diagonal matrices $ \mathcal{D}_{1,2}= \text{diag}(|d_{1,2}\rangle)$. Using this convention, we have
\begin{equation}
\mathcal{D}_2 = \frac{1}{\sqrt{3}}(C_3\mathcal{D}_1C_3^{-1}-C_3^{-1}\mathcal{D}_1C_3) , ~~
\text{Tr}(\mathcal{D}_{1}^2) = \text{Tr}(\mathcal{D}_{2}^2) = 1,~~\text{Tr}(\mathcal{D}_{1} \mathcal{D}_{2})  = 0, 
\label{eq:relationD12}
\end{equation}
and 
\begin{equation}
\text{Tr}(\mathcal{D}_{1} U \mathcal{D}_{1})  = \text{Tr}(\mathcal{D}_{2} U \mathcal{D}_{2}), ~~M^{-1} \mathcal{D}_{1,2}M = \pm \mathcal{D}_{1,2}.
\label{eq:theorem2}
\end{equation}
By substituting Eq.~(\ref{eq:linearDel}) into Eq.~(\ref{eq:freeenergyapp}), the free energy reads
\begin{equation}
\begin{aligned}
\mathcal{F} \approx & - \frac{1}{V_s^d} (|\Delta_1|^2+|\Delta_2|^2)  +   \text{Tr} \left\{ K_1  \left[ |\Delta_1|^2 \mathcal{D}_1^2    + |\Delta_2|^2  \mathcal{D}_2^2   
+2\text{Re}( \Delta_1 \Delta_2^* )   \mathcal{D}_1\mathcal{D}_2 \right]  \right\}   \\
 & +\frac{1}{2}\text{Tr} \left\{  K_2 \left[ |\Delta_1|^2 \mathcal{D}_1^2    + |\Delta_2|^2  \mathcal{D}_2^2   
+2\text{Re}( \Delta_1 \Delta_2^* )   \mathcal{D}_1\mathcal{D}_2 \right]^2 \right\}.  
 \end{aligned}
\end{equation}
Based on the definitions of $K_{1,2}$ given in Eq.~(\ref{eq:Kdefinition}), we have 
\begin{equation}
K_{1,2} = M^{-1} K_{1,2} M = C_3^{-1} K_{1,2} C_3.
\end{equation}
By combining with Eqs.~(\ref{eq:relationD12}) and (\ref{eq:theorem2}), 
\begin{equation}
\begin{aligned}
\text{Tr}(K_1 \mathcal{D}_1  \mathcal{D}_2) & = \text{Tr}(M^{-1} K_1M M^{-1} \mathcal{D}_1  M M^{-1}\mathcal{D}_2M) = -\text{Tr}(K_1 \mathcal{D}_1  \mathcal{D}_2)  = 0, \\
\text{Tr}(K_2 \mathcal{D}_1^3 \mathcal{D}_2) &= \text{Tr}(K_2 \mathcal{D}_1 \mathcal{D}_2^3   ) = 0,
\end{aligned}
\end{equation}
and 
\begin{equation}
\begin{aligned}
 \text{Tr}(K_1 \mathcal{D}_{1}^2)  = \text{Tr}(K_1 \mathcal{D}_{2}^2) , ~~~~\text{Tr}(K_2 \mathcal{D}_{1}^4)  = \text{Tr}(K_2 \mathcal{D}_{2}^4)= 3 \text{Tr}(K_2\mathcal{D}_{1}^2  \mathcal{D}_{2}^2)
\end{aligned}
\end{equation}
The last equality in the above equation can be proved as follows: For a general matrix $U$ that is invariant under $C_3$, we have
\begin{equation}
\text{Tr} (U \mathcal{D}_{1}) = \frac{1}{3}\text{Tr} [ (U +C_3^{-1}UC_3 ++C_3UC_3^{-1} )\mathcal{D}_{1}] = \frac{1}{3}\text{Tr} [U (\mathcal{D}_{1} + C_3^{-1} \mathcal{D}_{1} C_3 + C_3 \mathcal{D}_{1} C_3^{-1})]  = 0, 
\end{equation}
where we have used the relation $\mathcal{D}_{1} + C_3^{-1} \mathcal{D}_{1} C_3 + C_3 \mathcal{D}_{1} C_3^{-1}  = 0 $. Then we have
\begin{equation}
\text{Tr}(K_2\mathcal{D}_{1}^4) -3\text{Tr}(K_2\mathcal{D}_{1}^2 \mathcal{D}_{2}^2) = \text{Tr}[K_2\mathcal{D}_{1}^2(\mathcal{D}_{1}^2-3\mathcal{D}_{2}^2)] = 4\text{Tr}(K_2 \mathcal{D}_{1} C_3 \mathcal{D}_{1} C_3 \mathcal{D}_{1} C_3 \mathcal{D}_{1})  = 0
\end{equation}
because $C_{3} (K_2\mathcal{D}_{1} C_3 \mathcal{D}_{1} C_3 \mathcal{D}_{1} C_3) C_{3}^{-1} =  K_2\mathcal{D}_{1} C_3 \mathcal{D}_{1} C_3 \mathcal{D}_{1} C_3$ is invariant under $C_3$ operation. Based on these results, the free energy can be simplified as 
\begin{equation}
\begin{aligned}
\mathcal{F} = & -(\lambda_1+ \frac{1}{V_s^d}) (|\Delta_1|^2+|\Delta_2|^2)   + \frac{1}{2}\lambda_2 \left\{ (|\Delta_1|^2+|\Delta_2|^2)^2  -\frac{4}{3} |\Delta_1|^2|\Delta_2|^2 + \frac{4}{3}  [\text{Re}(\Delta_1\Delta_2^*)]^2\right\},  
 \end{aligned}
 \label{eq:free_energy}
\end{equation}
where 
\begin{equation}
\lambda_1 =  -\text{Tr}(K_1 \mathcal{D}_{1}^2)>0, ~~~~ \lambda_2 = \text{Tr}(K_2 \mathcal{D}_{1}^4)>0.
\end{equation}
In order to minimize the free energy, the second last term in Eq.~(\ref{eq:free_energy}) requires $|\Delta_1| = |\Delta_2|$, and the last term in Eq.~(\ref{eq:free_energy}) further requires the phase difference between $\Delta_1$ and $\Delta_2$ is $\pi/2$. Therefore, chiral $d_{x^2-y^2}\pm i d_{xy}$-wave pairing is energetically more favorable below $T_c$ for the spin-singlet channel. Since the above arguments depend only on the symmetries of the system, similar analysis can be applied to spin-triplet $p$-wave pairing, and chiral $p_{x}\pm i p_{y}$-wave pairing turns out to be energetically more favorable below $T_c$.

\section{Critical temperature and competing phases}
Quantitative prediction of critical temperature $T_c$ from FRG calculations stands as a challenging issue in the field not only because the computing time increases dramatically with calculation accuracy but also because the perturbation RG breaks down when the system flows out of the weak-coupling regime to the vicinity of an instability \cite{Metzner:2012aa}. Here we summarize three typical scenarios developed in earlier literature for estimating $T_c$. In the first scenario, $T_c$ is determined by the energy or temperature scale where the effective interaction possesses a pole \cite{Honerkamp:2001aa}. In numerical calculations, $T_c$ is approximated by the characteristic energy scale where the largest effective interaction exceeds a high value larger than the band width of the system. The second scenario is to explicitly calculate the RG flows of the susceptibilities \cite{Halboth:2000aa}, and $T_c$ is then identified as the energy scale where the susceptibility diverges. Commonly, the susceptibility of leading ordering tendency and the largest effective interaction diverge at the same energy scale \cite{Halboth:2000aa}. The third scenario of estimating $T_c$ also relies on identifying the pole of the susceptibility, which is instead calculated by performing a random phase approximation (RPA)-type resummation within an effective interaction model derived from RG calculations \cite{Koshino:2018aa,Chubukov:2008aa}. In the present work, we estimate $T_c$ from solving the linearized gap equation within an effective interaction model, as detailed in the following subsection.

\subsection{Pairing channel}
\label{subsec:pairing}
The temperature-flow FRG calculations result in effective interactions between electrons at different temperature scales. Here we choose the pairing channel as an example to show the scheme employed in this study for estimating $T_c$. By focusing on pairing interaction between electrons from opposite valleys, the effective interaction model is  
\begin{equation}
H_{eff} = \sum_{\tau n } \sum_{\bm{k} \in n}\epsilon_{ \tau n}(\bm{k}) \psi^{\dagger}_{\tau n }  \psi_{\tau n } + \sum_{\tau  n \tau_1 n_1} u( \tau n, \bar{\tau} \bar{n}; \bar{\tau}_1 \bar{n}_1,\tau_1 n_1) \psi^{\dagger}_{\tau n } \psi^{\dagger}_{\bar{\tau} \bar{n} }  \psi_{\bar{\tau}_1\bar{n}_1 } \psi_{\tau_1 n_1 },
\end{equation}
where $\epsilon_{ \tau n}(\bm{k})$ is the valley- and patch-resolved electron energy at wave vector $\bm{k}$. Within mean-field theory, the pairing potential at patch $n$ is 
\begin{equation}
\Delta(\tau n,\bar{\tau} \bar{n}) = \sum_{ \tau_1n_1}  u( \tau n, \bar{\tau} \bar{n}; \bar{\tau}_1 \bar{n}_1,\tau_1 n_1)  \langle  \psi_{\bar{\tau}_1\bar{n}_1 } \psi_{\tau_1 n_1 } \rangle.
\end{equation}
The corresponding effective action 
\begin{equation}
S_{eff}[\psi^{\dagger}, \psi] =  -\sum_{\tau n } \int_{P} \psi^{\dagger}_{\tau n ,P} (i\omega-\epsilon_{ \tau n,\bm{k}}) \psi_{\tau n ,P} +
 \sum_{\tau n}  \int_{P}
 \left[\Delta (\tau n,\bar{\tau} \bar{n}) \psi^{\dagger}_{\tau n ,P}  \psi^{\dagger}_{\bar{\tau} \bar{n} ,-P}  +\text{H.c.} \right] -  \Delta^{\dagger} u^{-1} \Delta 
\end{equation}
where $P = (i\omega,\bm{k})$, definition of $\int_{P} $ is given by Eq.~(\ref{eq:intdefine}) in Sec.~\ref{subsec:Effect}, $u^{-1}$ denotes the inverse matrix of the pairing interaction matrix, and $\Delta = [\cdots, \Delta (\tau n,\bar{\tau} \bar{n}) ,\cdots]^{\text{T}}$ is a vector notation of the pairing potential.
Similar to the derivations in Sec.~\ref{subsec:Effect}, by integrating out the fermionic fields and expanding the free energy up to quartic order of $\Delta$, we have an effective Landau free energy 
\begin{equation}
\mathcal{F} = - \Delta^{\dagger} u^{-1} \Delta +  \Delta^{\dagger} K_1\Delta   +\frac{1}{2} \Delta^{\dagger} \Delta^{\dagger} K_2 \Delta \Delta.
\label{eq:free_energy_SC}
\end{equation}
Here $K_{1,2}$ are patch- and valley-diagonal matrices defined as
\begin{equation}
\begin{aligned}
K_1(\tau n) &= k_BT\sum_{i\omega} \sum_{\bm{k} \in n} \frac{1}{ (i\omega +   \epsilon_{\tau n,\bm{k} })(i\omega -   \epsilon_{\tau n,\bm{k} })}=-\sum_{\bm{k} \in n} \frac{1-2f(\epsilon_{\tau n,\bm{k}})}{2\epsilon_{\tau n,\bm{k}}} =- \Pi^{pp} (\tau n, 0) , \\
K_2 (\tau n)& =  k_BT\sum_{i\omega} \sum_{\bm{k} \in n} \frac{1}{ (i\omega +   \epsilon_{\tau n,\bm{k} })^2(i\omega -   \epsilon_{\tau n,\bm{k} })^2} .
\end{aligned}
\label{eq:Kdefinition}
\end{equation}
In the above equations, $\Pi^{pp}(\tau n, 0) $ are patch-resolved zero-$\bm{q}$ non-interaction static particle-particle susceptibilities.
Pairing instability sets in when the coefficient of the quadratic terms in Eq.~(\ref{eq:free_energy_SC}) becomes negative. Therefore, $T_c$ is determined by 
\begin{equation}
\Delta^{\dagger} (K_1 - u^{-1}) \Delta =0,
\end{equation}
which reduces to the following linearized gap equation 
\begin{equation}
 \Delta = u K_1 \Delta. 
 \label{eq:LGE}
\end{equation}

\begin{figure}
 \centering
\includegraphics[width=0.9\columnwidth]{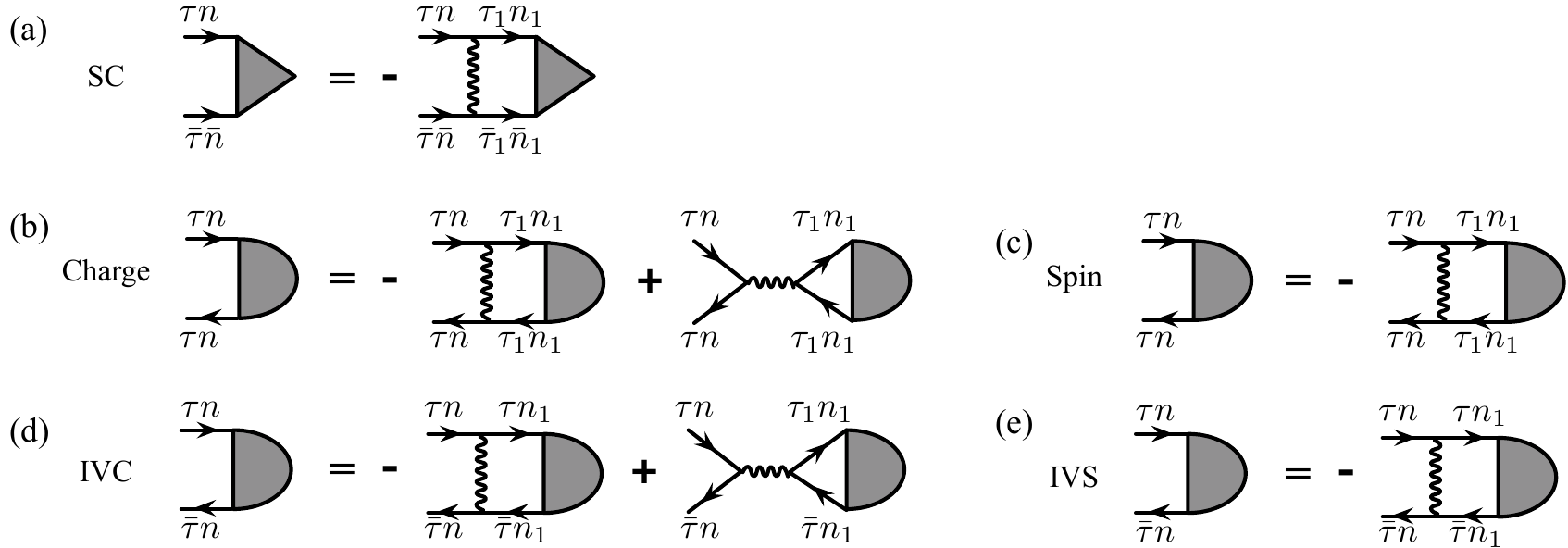}
 \caption{Diagrammatic representation of the linearized gap equations for (a) pairing channel, (b) charge order channel, (c) spin order channel, (d) inter-valley charge coherence, and (e) inter-valley spin coherence, where the wavy lines represent renormalized 4PVs. Here the filled triangle and semicircle denote order parameters of the particle-particle (pairing) and particle-hole channels, respectively. The valley conservation has been employed in (d) and (e).}
 \label{fig:figureS4}
\end{figure}

Diagrammatic representation of Eq.~(\ref{eq:LGE}) is shown in Fig.~\ref{fig:figureS4}(a), which reads as
\begin{equation}
\Delta( \tau n, \bar{\tau} \bar{n}) = -\sum_{\tau_1 n_1} \Pi^{pp}(\tau_1n_1,0) u( \tau_1 n_1, \bar{\tau}_1 \bar{n}_1; \bar{\tau} \bar{n},\tau n) \Delta(\tau_1 n_1, \bar{\tau}_1 \bar{n}_1).
\end{equation}
By inserting Eqs.~(\ref{eq:time-reversal}) and (\ref{eq:classify}) in the above equation, 
\begin{equation}
\begin{aligned}
\Delta(Kn,  K'\bar{n}) &= -\sum_{n_1}  \Pi^{pp}_{KK} (n_1,0) [u_t(n_1 ,\bar{n}_1 ;\bar{n} ,n ) \Delta(Kn_1,K'\bar{n}_1 ) 
+ u_e(n_1 ,\bar{n}_1 ;n ,\bar{n} ) \Delta(K'\bar{n}_1,Kn_1 )], \\
\Delta( K'\bar{n},Kn) &= -\sum_{n_1}  \Pi^{pp}_{KK} (n_1,0) [u_e(n_1 ,\bar{n}_1 ;n,\bar{n} ) \Delta(Kn_1,K'\bar{n}_1 ) 
+ u_t(n_1 ,\bar{n}_1 ; \bar{n},n ) \Delta(K'\bar{n}_1,Kn_1 )].
\end{aligned}
\end{equation}
The pairing potential can be organized into even-parity spin-singlet ($\Delta_{s}$) and odd-parity spin-triplet ($\Delta_{t}$) channels as
\begin{equation}
\begin{aligned}
\Delta_{s}(n) &=  \Delta(Kn, K'\bar{n} )  +  \Delta( K'\bar{n},Kn), \\
\Delta_{t}(n) &=  \Delta(Kn,K' \bar{n} )  -  \Delta( K'\bar{n} ,Kn).
\end{aligned}
\end{equation}
The corresponding linearized gap equations for spin-singlet and spin-triplet channels are 
\begin{equation}
\begin{aligned}
\Delta_{s}(n) &=  -\sum_{n_1}  \Pi^{pp}_{KK} (n_1,0) [ u_t(n_1 ,\bar{n}_1 ; \bar{n},n ) +u_e(n_1 ,\bar{n}_1 ;n,\bar{n} ) ] \Delta_{s}(n_1) = -\sum_{n_1}  \Pi^{pp}_{KK} (n_1,0) V_s(n_1;n) \Delta_{s}(n_1) , \\
\Delta_{t}(n) &=  -\sum_{n_1} \Pi^{pp}_{KK} (n_1,0) [ u_t(n_1 ,\bar{n}_1 ; \bar{n},n ) -u_e(n_1 ,\bar{n}_1 ;n,\bar{n} ) ] \Delta_{t}(n_1) =-\sum_{n_1}  \Pi^{pp}_{KK} (n_1,0) V_t(n_1;n) \Delta_{t}(n_1) , 
\end{aligned}
\label{eq:linearizedgap}
\end{equation}
which in turn justify the definitions of spin-singlet and spin-triplet pairing interactions given in Eq.~(3) in the main text. Since both $\Pi^{pp}_{KK}$ and $V_{s,t}$ are temperature dependent, $T_c's$ for different pairing channels are determined by the  temperatures where the corresponding eigenvalues of $-\Pi^{pp}_{KK}V_{s,t}$ equal 1. Specifically, Eq.~(\ref{eq:linearizedgap}) indicates repulsive (attractive) valley-exchange interaction prefers spin-triplet (singlet) pairing.

\begin{figure}
 \centering
\includegraphics[width=0.4\columnwidth]{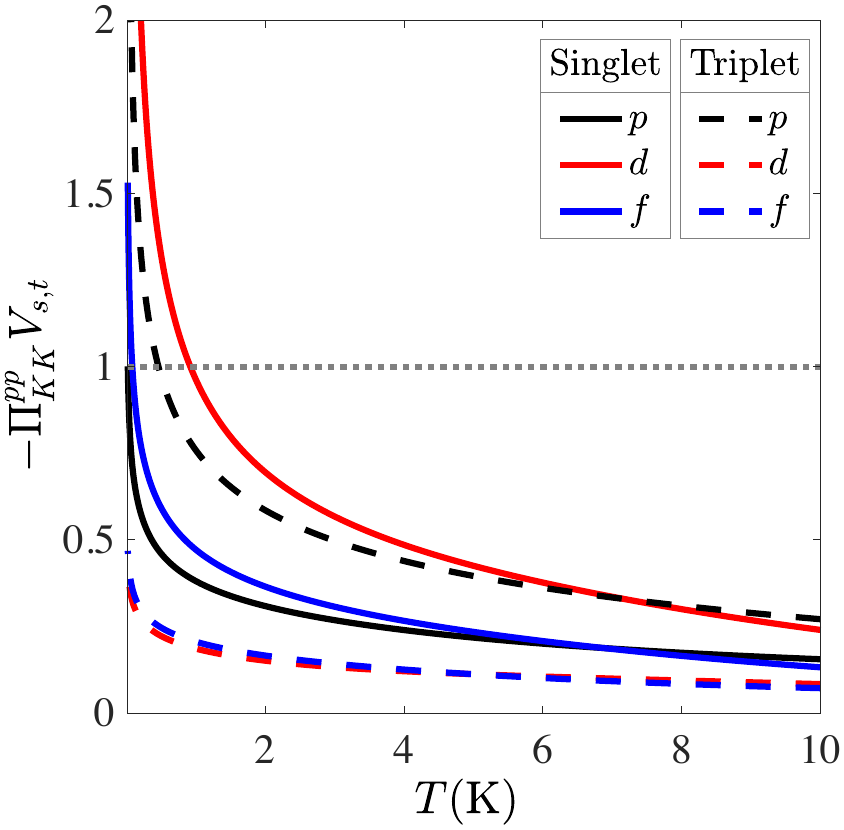}
 \caption{Eigenvalues of $-\Pi^{pp}_{KK}V_{s,t}$ versus temperature, where $T_c's$ are estimated by the temperatures of intersections between these eigenvalues and 1 (dotted line). These calculations are obtained by choosing $\epsilon=5$, $d_s=40$ nm, $\Delta_d = 30$ meV, and $n_e = -1.75\times10^{12} $ cm$^{-2}$.}
 \label{fig:figureR1}
\end{figure}

Figure \ref{fig:figureR1} shows temperature dependences of the eigenvalues of $-\Pi^{pp}_{KK}V_{s,t}$, where $T_c's$ are determined by the intersections between them and the dotted line. The linearized gap equation scheme given above is indeed equivalent to the third scenario of estimating $T_c$, where the dressed susceptibility $\chi_{\eta}$ is obtained from a RPA-type resummation \cite{Koshino:2018aa,Chubukov:2008aa}
\begin{equation}
\chi_{\eta} = \frac{\chi_{\eta}^0}{1+V_{\eta}\chi_{\eta}^0},
\end{equation}
where $\chi_{\eta}^0$ is the bare susceptibility of ordering channel $\eta$ and $V_{\eta}$ denotes the corresponding effective interaction obtained from RG calculations. Divergence of $\chi_{\eta}$ is then characterized by $V_{\eta}\chi_{\eta}^0=-1$, recovering the condition derived from the linearized gap equation shown in Eq.~(\ref{eq:linearizedgap}). 
We note that identifying the pole of the effective pairing interaction shown in Fig.~3(a) in the main text is not equivalent to solving the linearized gap equation in estimating $T_c$. The linearized gap equation scheme employed in this study (or RPA-type resummation of susceptibility \cite{Koshino:2018aa,Chubukov:2008aa}) has the shortcoming of double counting a subsect of Feynman diagrams, which may alter the quantitative values of $T_c$. Nevertheless, we emphasize that the qualitative features, such as the leading pairing instability and pairing symmetry, are almost unaffected. For example, as shown in Fig.~3(a) in the main text and Fig.~\ref{fig:figureR1}, the results obtained from identifying the pole of the effective pairing interaction (see Fig.~3(a)) and from solving the linearized gap equation (see Fig.~\ref{fig:figureR1}) show that the leading instability is spin-singlet $d$-wave pairing. Meanwhile, $T_c$ estimated from linearized gap equation is slightly higher than that estimated from the pole of effective pairing interaction.

\subsection{Particle-hole channels}
\label{subsec:particle-hole}
As illustrated in Fig.~2 in the main text, the valley-exchange interaction ($u_e$) is strengthened under RG and flows to be comparable in magnitude with the inter-valley interaction ($u_t$). The presence of $u_e$ breaks the valley-SU$(2)$ symmetry, lifting the degeneracy between spin-singlet (valley-triplet) and spin-triplet (valley-singlet) pairing. Since the spin-orbit coupling (SOC) is weak in RTG, we invoke the spin-SU($2$) symmetry by ignoring the SOC effect. We focus on several typical particle-hole channel instabilities summarized in Table. I in the main text. Similar to the pairing instability, the critical temperatures of these particle-hole instabilities can be estimated via their corresponding linearized gap equations, as depicted graphically in Fig.~\ref{fig:figureS4}.

\subsubsection{Charge order}
Figure \ref{fig:figureS4}(b) shows the diagrammatic representation of the linearized gap equation for charge order channel, which reads
\begin{equation}
 \Delta(\tau n, \tau n) = -\sum_{\tau_1 n_1} \Pi^{ph}(\tau_1 n_1,0) 
[u(\tau_1n_1,\tau n ;\tau_1n_1,\tau n ) -2 u(\tau_1n_1,\tau n ;\tau n,\tau_1n_1 )  ] \Delta(\tau_1n_1,\tau_1n_1),
\end{equation}
where $ \Delta(\tau n, \tau n) = \Delta(\uparrow\tau n , \uparrow \tau n ) + \Delta(\downarrow\tau n , \downarrow \tau n ) $ denotes the charge order parameter at valley $\tau$ and patch $n$. By inserting Eqs.~(\ref{eq:time-reversal}) and (\ref{eq:classify}) into the above gap equation,
\begin{equation}
\begin{aligned}
\Delta(Kn,Kn) &= -\sum_{n_1} \Pi^{ph}_{KK}(n_1,0) 
[u_a(n_1,n;n_1,n) -2 u_a(n_1,n;n,n_1)] \Delta(Kn_1,Kn_1) \\
&~~~ -\sum_{n_1}  \Pi^{ph}_{KK}(n_1,0) 
[u_e(n_1,\bar{n};n_1,\bar{n}) -2 u_t(n_1,\bar{n};\bar{n},n_1)] \Delta(K'\bar{n}_1,K'\bar{n}_1), \\
\Delta(K'\bar{n},K'\bar{n}) &= -\sum_{n_1}  \Pi^{ph}_{KK}(n_1,0) 
[u_e(n_1,\bar{n};n_1,\bar{n}) -2 u_t(n_1,\bar{n};\bar{n},n_1)] \Delta(Kn_1,Kn_1) \\
&~~~ -\sum_{n_1}  \Pi^{ph}_{KK}(n_1,0) 
[u_a(n_1,n;n_1,n) -2 u_a(n_1,n;n,n_1)] \Delta(K'\bar{n}_1,K'\bar{n}_1). \\
\end{aligned}
\end{equation}
By further organizing the order parameter into even-parity ($\Delta \propto s_0\tau_0$) and odd-parity  ($\Delta \propto s_0\tau_z$) channels (see Table. I in the main text), we have 
\begin{equation}
\begin{aligned}
\Delta_{s_0\tau_0}(n) &= \Delta(Kn,Kn) + \Delta(K'\bar{n},K'\bar{n}), \\
\Delta_{s_0\tau_z}(n) &= \Delta(Kn,Kn) - \Delta(K'\bar{n},K'\bar{n}),
\end{aligned}
\end{equation}
and
\begin{equation}
\begin{aligned}
\Delta_{s_0\tau_0}(n) &= -\sum_{n_1} \Pi^{ph}_{KK}(n_1,0) 
[u_a(n_1,n;n_1,n) -2 u_a(n_1,n;n,n_1) + u_e(n_1,\bar{n};n_1,\bar{n}) -2 u_t(n_1,\bar{n};\bar{n},n_1)] \Delta_{s_0\tau_0}(n_1), \\
\Delta_{s_0\tau_z}(n) &= -\sum_{n_1} \Pi^{ph}_{KK}(n_1,0) 
[u_a(n_1,n;n_1,n) -2 u_a(n_1,n;n,n_1) - u_e(n_1,\bar{n};n_1,\bar{n}) +2 u_t(n_1,\bar{n};\bar{n},n_1)] \Delta_{s_0\tau_z}(n_1).
\end{aligned}
\end{equation}
Here the even-parity order parameter $\Delta_{s_0\tau_0}$ represents to Pomeranchuk instability (PI), arising from the deformation of the shape of the Fermi surface. The odd-parity order parameter $\Delta_{s_0\tau_0}$ denotes valley polarization (VP).

\subsubsection{Spin order}
Particle-hole instability in the spin order channel arises from exchange-type interaction. Figure \ref{fig:figureS4}(c) depicts the diagram of the linearized gap equation for the spin order, which reads 
\begin{equation}
 \Delta(\tau n, \tau n) = -\sum_{\tau_1 n_1} \Pi^{ph}(\tau_1 n_1,0) 
u(\tau_1n_1,\tau n ;\tau_1n_1,\tau n )  \Delta(\tau_1n_1,\tau_1n_1),
\end{equation}
where $ \Delta(\tau n, \tau n) = \Delta(\uparrow\tau n , \uparrow \tau n ) - \Delta(\downarrow\tau n , \downarrow \tau n ) $ denotes the spin order parameter at valley $\tau$ and patch $n$. By inserting Eqs.~(\ref{eq:time-reversal}) and (\ref{eq:classify}) into the above gap equation, 
\begin{equation}
\begin{aligned}
\Delta(Kn,Kn) &= -\sum_{n_1} \Pi^{ph}_{KK}(n_1,0) 
[u_a(n_1,n;n_1,n) \Delta(Kn_1,Kn_1) +
u_e(n_1,\bar{n};n_1,\bar{n})  \Delta(K'\bar{n}_1,K'\bar{n}_1)], \\
\Delta(K'\bar{n},K'\bar{n}) &= -\sum_{n_1}  \Pi^{ph}_{KK}(n_1,0) 
[u_e(n_1,\bar{n};n_1,\bar{n})  \Delta(Kn_1,Kn_1) + 
u_a(n_1,n;n_1,n)  \Delta(K'\bar{n}_1,K'\bar{n}_1)]. \\
\end{aligned}
\end{equation}
Similarly, the order parameter can be organized into even-parity ($\Delta \propto s_z\tau_0$) and odd-parity ($\Delta \propto s_z\tau_z$) channels, 
\begin{equation}
\begin{aligned}
\Delta_{s_z\tau_0}(n) &= \Delta(Kn,Kn) + \Delta(K'\bar{n},K'\bar{n}), \\
\Delta_{s_z\tau_z}(n) &= \Delta(Kn,Kn) - \Delta(K'\bar{n},K'\bar{n}),
\end{aligned}
\end{equation}
and
\begin{equation}
\begin{aligned}
\Delta_{s_z\tau_0}(n) &= -\sum_{n_1} \Pi^{ph}_{KK}(n_1,0) 
[u_a(n_1,n;n_1,n)  + u_e(n_1,\bar{n};n_1,\bar{n}) ] \Delta_{s_z\tau_0}(n_1), \\
\Delta_{s_z\tau_z}(n) &= -\sum_{n_1} \Pi^{ph}_{KK}(n_1,0) 
[u_a(n_1,n;n_1,n)  - u_e(n_1,\bar{n};n_1,\bar{n}) ] \Delta_{s_z\tau_z}(n_1).
\end{aligned}
\label{eq:linearspin}
\end{equation}
Here $\Delta_{s_z\tau_0}$ and $\Delta_{s_z\tau_z}$ are order parameters of the valley-unpolarized ferromagnetism (FM) and anti-ferromagnetism (AFM), which are distinguished by identical and opposite spin polarizations in $K$ and $K'$ valleys. In particular, Eq.~(\ref{eq:linearspin}) suggests the degeneracy between FM and AFM is lifted by valley-exchange interaction $u_e$, with repulsive (attractive) $u_e$ preferring FM (AFM).

\subsubsection{Inter-valley coherence}
As summarized in Table. I in the main text, the inter-valley coherence states are classified into inter-valley charge (IVC, $\Delta \propto s_0\tau_{x,y}$) and  inter-valley spin (IVS, $\Delta \propto s_z\tau_{x,y}$) states. As shown in Fig.~\ref{fig:figureS4}(d), the linearized gap equation for IVC reads
\begin{equation}
 \Delta(\tau n, \bar{\tau} n) = -\sum_{\tau_1 n_1} \Pi^{ph}(\tau_1 n_1,2\bm{K}) 
[u(\bar{\tau}_1n_1,\tau n ;\tau_1n_1,\bar{\tau} n ) -2 u(\bar{\tau}_1n_1,\tau n ;\bar{\tau} n,\tau_1n_1 )  ] \Delta(\tau_1n_1,\bar{\tau}_1n_1).
\end{equation}
By further insisting the valley conservation $\tau + \bar{\tau}_1= \tau_1 + \bar{\tau}$, we have $\tau_1 = \tau$ and
\begin{equation}
 \Delta_{s_0\tau_{xy}}(n) = -\sum_{n_1} \Pi^{ph}_{KK'}( n_1,0) 
[u_t^*(n_1,n ;n_1, n) -2 u_e^*(n_1,n ;n,n_1 )  ] \Delta_{s_0\tau_{xy}}(n_1) ,
\end{equation}
where relations given in Eqs.~(\ref{eq:time-reversal}) and (\ref{eq:classify}) are employed and $\Pi^{ph}_{KK'} $ denotes the inter-valley particle-hole susceptibility. Only the exchange interaction contributes to the linearized gap equation for IVS, leading to 
\begin{equation}
 \Delta_{s_z\tau_{xy}}(n) = -\sum_{n_1} \Pi^{ph}_{KK'}( n_1,0) 
u_t^*(n_1,n ;n_1, n)  \Delta_{s_z\tau_{xy}}(n_1).
\label{eq:linIVS}
\end{equation}
The above equations suggest that the degeneracy between IVC and IVS is also lifted by $u_e$, with repulsive (attractive) $u_e$ preferring IVS (IVC).

\subsection{Phase diagram}
\begin{figure}
 \centering
\includegraphics[width=0.7\columnwidth]{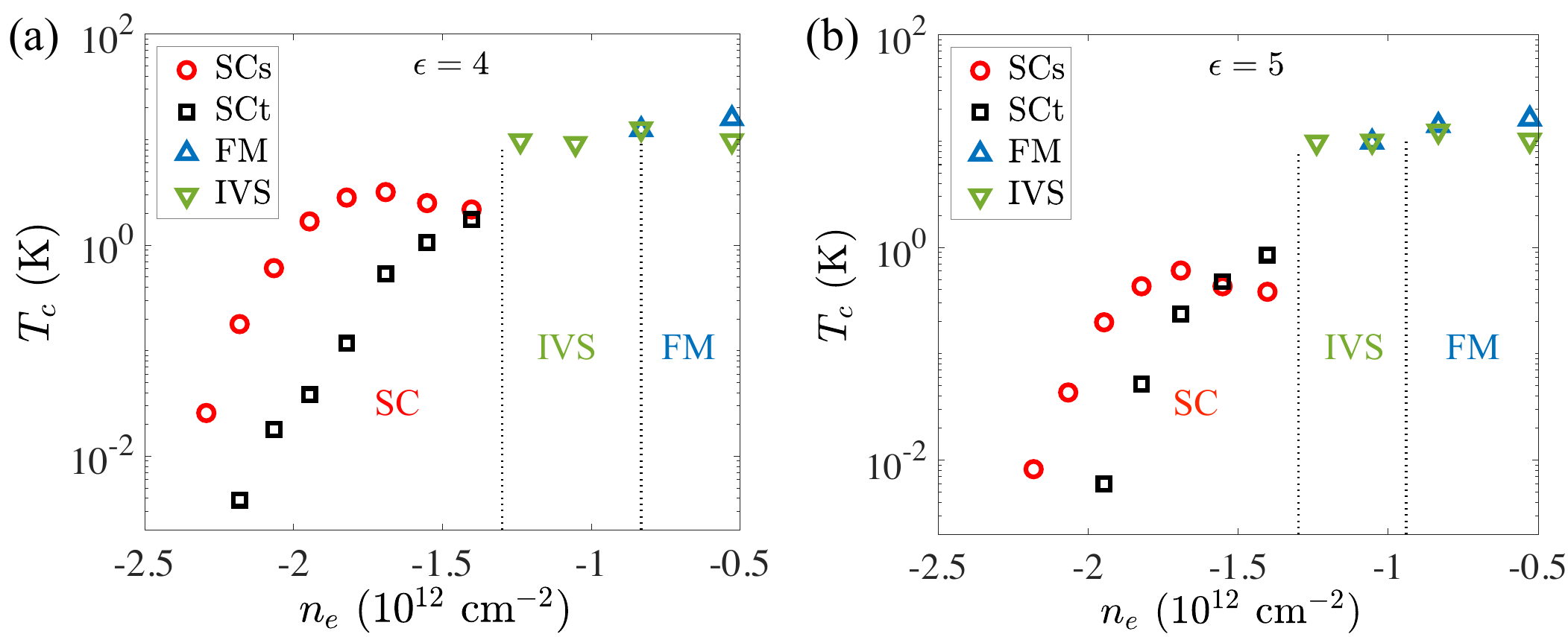}
 \caption{Phase diagrams as function of electron density $n_e$ for (a) $\epsilon=4$ and (b) $\epsilon=5$. Here SCs and SCt represent spin-singlet $d$-wave pairing and spin-triplet $p$-wave paring, respectively. These results are obtained by choosing $\Delta_d =30$ meV, $d_s = 40$ nm, and $u_{e,0} = v(Q) = 2\pi e^2/\epsilon Q$ with $Q = 2|\bm{K}|$ being the magnitude of the wavevector difference between two valleys. Dotted lines highlight the phase boundaries.}
 \label{fig:figureS5}
\end{figure}

By solving the linearized gap equations discussed in the above two subsections, we explore the competition between instabilities in pairing and particle-hole channels by varying model parameters, such as the displacement field-induced electrostatic potential $\Delta_d$, carrier density $n_e$, bare value of the valley-exchange interaction $u_{e,0}$, dielectric constant $\epsilon$, and gate-sample distance $d_s$. Figure~\ref{fig:figureS5} compares the phase diagrams as function of $n_e$ obtained by choosing two different values of $\epsilon$. Larger $\epsilon$ or weaker bare Coulomb interaction results in lower superconducting critical temperatures $T_c's$ for both of the spin-singlet and spin-triplet pairing channels. Particularly, as illustrated in Fig.~\ref{fig:figureS5}, $T_c$ for $d$-wave spin-singlet pairing is more sensitive to $\epsilon$ than that for spin-triplet $p$-wave pairing.

\begin{figure}
 \centering
\includegraphics[width=0.7\columnwidth]{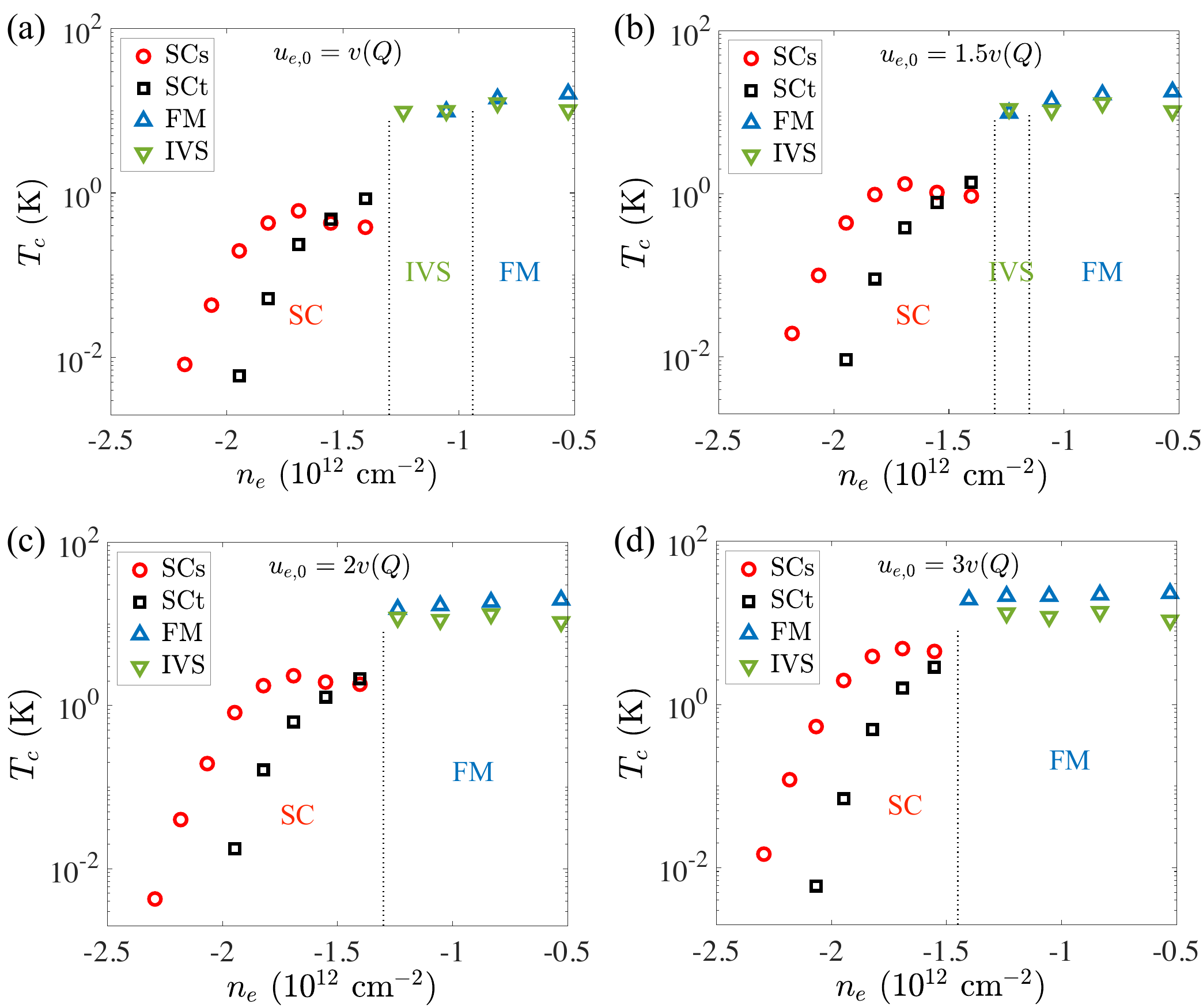}
 \caption{Phase diagrams as function of $n_e$ for several typical values of $u_{e,0}$. Here the definitions of notations are identical to that in Fig.~\ref{fig:figureS5}.}
 \label{fig:figureS6}
\end{figure}

As discussed in Sec.~\ref{sec:Coulomb}, $u_{e,0}$ is much smaller than the bare values of $u_{a,t}$. In fact, it is hard to give an accurate estimation of $u_{e,0}$ due to, for example, the presence of reciprocal lattice vectors. Hence, we explore the effect of varying $u_{e,0}$ on the phase diagram. The results of choosing several typical values of $u_{e,0}$ are summarized in Fig.~\ref{fig:figureS6}, where we find that the $T_c's$ for both of spin-singlet $d$-wave and spin-triplet $p$-wave pairings are enhanced upon increasing $u_{e,0}$. The spin-singlet $d$-wave pairing dominates over the spin-triplet $p$-wave pairing across the whole SC phase regime when $u_{e,0} $ is large enough, e.g. $u_{e,0} \geqslant 3v(Q)$, as shown in Fig.~\ref{fig:figureS6}(d).  In addition, larger values of $u_{e,0}$ prefer the FM state than the IVS state. This is because repulsive $u_{e}$ enhances the $T_c$ for the FM state as explained in Eq.~(\ref{eq:linearspin}), while $u_{e}$ is irrelevant to the linearized gap equation (Eq.~(\ref{eq:linIVS})) for the IVS state.


The phases diagrams depicted in Figs.~\ref{fig:figureS5} and \ref{fig:figureS6} possess similar qualitative features to the phase diagram shown in Fig.~5(a) in the main text. A general trend of superconducting $T_c$ is that larger bare Coulomb interaction leads to higher $T_c$, contradictory to the electron-phonon mechanism of superconductivity \cite{Chou:2021aa}. As discussed in the main text, the experimentally identified partially isospin polarized (PIP) state is consistent with the theoretically predicted IVS state. By expanding the IVS (PIP) state stability region to agree with experiment, the PIP state intervenes before the peak of the superconducting dome is reached and the triplet pairing state is never stabilized -both in agreement with experiment.  Such a shift will avoid the regions close to the IVS phase on the SC phase side, where the spin-triplet pairing state becomes nearly degenerate with or even dominates over the spin-singlet pairing.  In fact, a more accurate determination of the phase boundary is very demanding because, for example, the influence of self-energy renormalization on the band structure has been ignored in this study. In the large hole density regime, the self-energy effect is likely to be weak due to low density of states. By tuning the Fermi level close to the VHS, the self-energy effect becomes important and therefore may shift the phase boundary to a considerable degree. 

\begin{figure}
 \centering
\includegraphics[width=0.35\columnwidth]{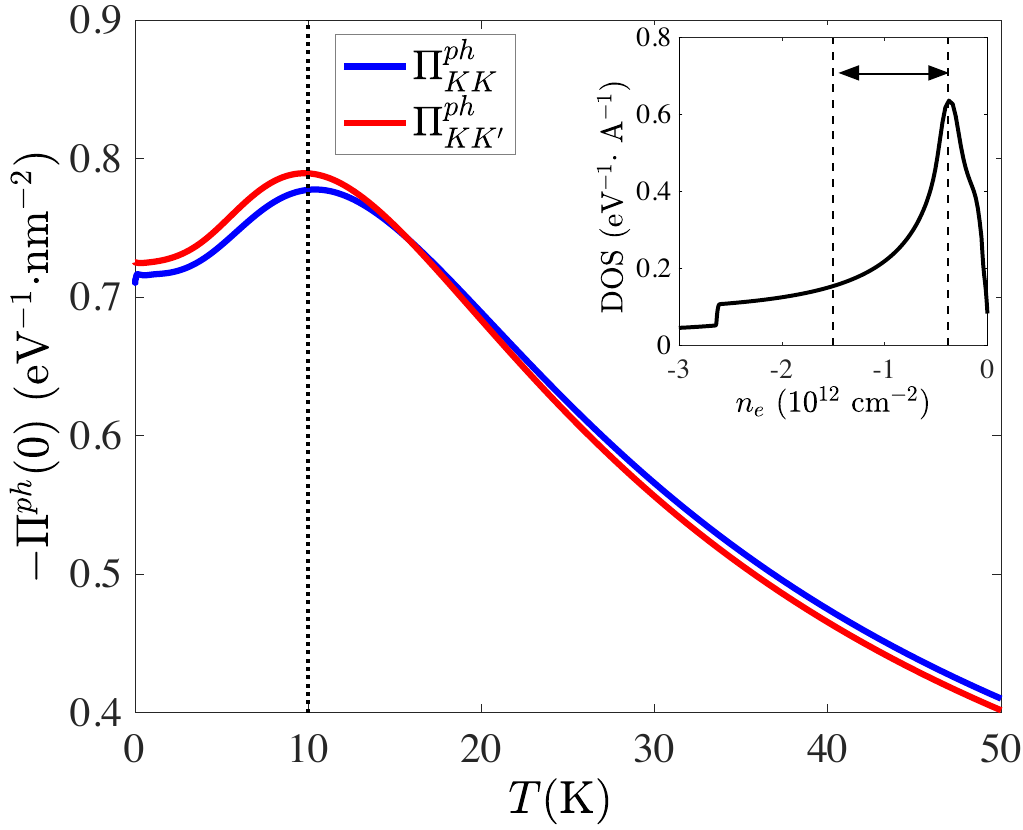}
 \caption{Temperature dependences of the zero-$q$ intra-valley ($\Pi_{KK}^{ph}$) and inter-valley ($\Pi_{KK'}^{ph}$) particle-hole susceptibilities calculated by choosing $n_e =-1.75\times 10^{12}$ cm$^{-2}$ and $\Delta_d=30$ meV. The dotted line marks the temperature where $\Pi_{KK}^{ph}$ and $\Pi_{KK'}^{ph}$ possess maximum values. The insert plots the density of states versus $n_e$.}
 \label{fig:figureS7}
\end{figure}

As pointed out in the main text, another feature of these phase diagrams is that $T_c's$ for the IVS and FM states exhibit sudden jumps near their phase boundaries. This phenomenon arises from the non-monotonic temperature dependence of the zero-$q$ intra-valley and inter-valley particle-hole susceptibilities, as plotted in Fig.~\ref{fig:figureS7}, which possess maximum values at finite temperatures due to proximity to the VHS (see inert). For example, $T_c$ for the IVS state is determined by Eq.~(\ref{eq:linIVS}), which turns out to first have finite-temperature solutions upon decreasing hole density and moving the Fermi level toward the VHS. Similar argument can be applied to the $T_c$ for the FM state. Moreover, as shown in Fig.~\ref{fig:figureS7}, the maximum value of $\Pi_{KK'}^{ph}$ is larger than that of $\Pi_{KK}^{ph}$, which is one of the main reasons that the IVS state emerges earlier than the FM state with increasing $n_e$, as indicated in Fig.~\ref{fig:figureS5}.

\begin{figure}
 \centering
\includegraphics[width=0.7\columnwidth]{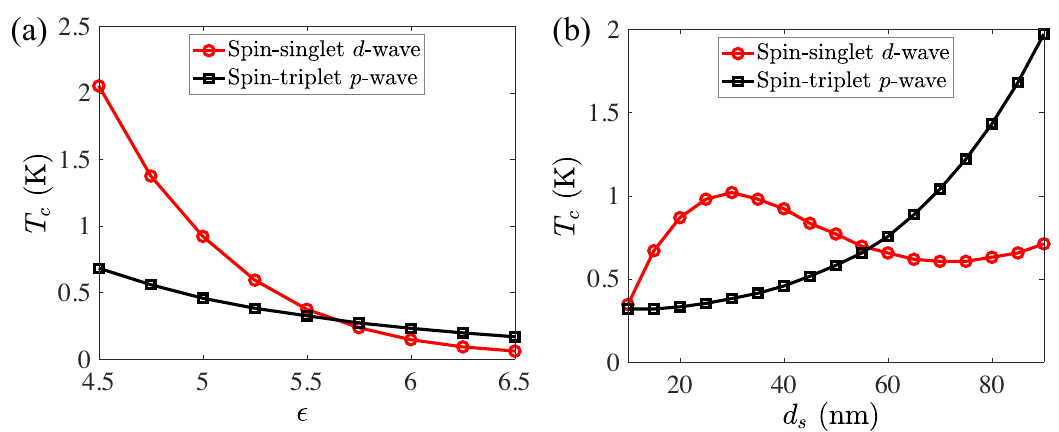}
 \caption{Superconducting $T_c$ versus (a) $\epsilon$ for $d_s = 40$ nm and (b) $d_s$ for $\epsilon =5$. These results are calculated by choosing $\Delta_d =30$ meV, $n_e =-1.75\times 10^{12}$ cm$^{-2}$, and $u_{e,0} = v(Q)$.}
 \label{fig:figureS8}
\end{figure}

Figure~\ref{fig:figureS8} explores the effects of altering background screening on superconducting $T_c$. Fig.~\ref{fig:figureS8}(a) shows that the $T_c's$ for both of the spin-singlet and spin-triplet 
channels are enhanced upon decreasing $\epsilon$, and a phase transition from spin-triplet to spin-singlet pairing can occur. Figure~\ref{fig:figureS8}(b) predicts that 
the distance to the metallic gates $d_s$ is also effective in altering $T_c$.
In this case, the spin-singlet channel has dome-like behavior,
possibly associated with a change in the delicate competition 
between scattering processes on the inner and outer Fermi surfaces 
since gate screening is momentum dependent. The dome peak occurs at $d_s \sim 30$ nm, 
near the onset of gate screenting at the inter-Fermi surface nesting wavevector.

\section{Gap functions and pairing interactions}
\subsection{Gap functions}
\begin{figure}
 \centering
\includegraphics[width=0.99\columnwidth]{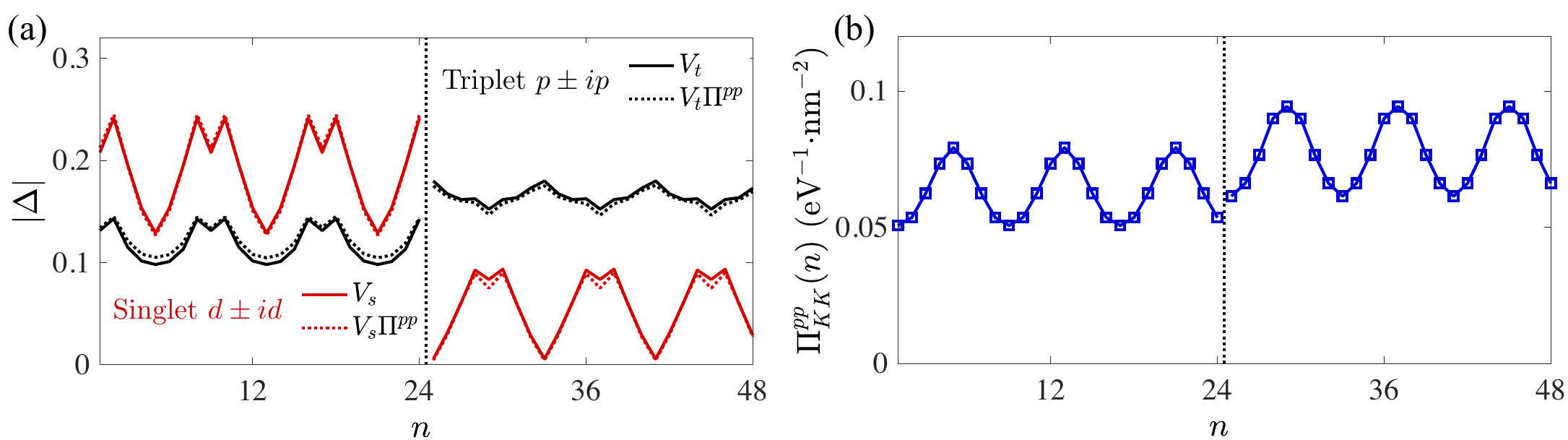}
 \caption{(a) Magnitudes of spin-singlet $d\pm id$ (red) and spin-triplet $p\pm ip$ (black) gap functions on the inner and outer Fermi surfaces. The solid lines depict the eigenvectors of pairing interactions $V_{s,t}$. The dotted lines are obtained by solving the corresponding linearized gap equations, namely, the eigenvectors of $V_{s,t} \Pi^{pp}$. These results are calculated at critical temperatures $T_{c,d}$ and $T_{c,p}$ for spin-singlet $d$-wave-like and spin-triplet $p$-wave-like channels, respectively. (b) Non-interacting static particle-particle susceptibility calculated at $T_{c,d}$.}
 \label{fig:figureS9}
\end{figure}

The gap function at superconducting critical temperature is in general given by the solution of the linearized gap equation. Within the patch scheme employed in the present study, the linearized gap equation is depicted graphically in Fig.~\ref{fig:figureS4}(a) and given explicitly by Eq.~(\ref{eq:linearizedgap}). Solving the linearized gap equation amounts to finding the eigenvalues and eigenvectors of matrix $V_{s,t} \Pi^{pp}$, where $V_{s,t}$ are the spin-singlet and spin-triple pairing interactions, and diagonal matrix $\Pi^{pp} = \text{diag}[\Pi^{pp}_{KK}(n)]$ defines the patch-specified zero-$q$ particle-particle susceptibility. The dominant pairing symmetry is characterized by the eigenvectors of the largest magnitude negative eigenvalue of $V_{s,t} \Pi^{pp}$ at $T_c$. Here we show that the gap functions are mainly determined by the matrix structures of $V_{s,t}$ instead of $\Pi^{pp}$, and therefore can be well approximated by the eigenvectors of $V_{s,t}$. As illustrated in Fig.~\ref{fig:figureS9}(a), the difference between gap functions $\Delta_{s,t}$ obtained by solving linearized the gap equations and the eigenvectors of the smallest eigenvalues of $V_{s,t}$ are negligible. Therefore, in the main text, we use the eigenvectors of $V_{s,t}$ to approximate $\Delta_{s,t}$.

Taking spin-singlet pairing as an example, we next demonstrate the numerical finding that $\Delta_{s}$ is mainly determined by the matrix structure of $V_{s}$. Because $V_{s}$ is invariant under $C_3$ symmetry, the 2D irreducible representation (IR) of which has two-degenerate eigenvectors. As shown in Sec.~\ref{subsec:chiral_pairing}, the smallest eigenvalue of $V_{s}$ is $V_{s}^{d}$, characterized by the following equation
\begin{equation}
V_{s} |d_{\pm} \rangle = V_{s}^{d}  |d_{\pm} \rangle,
\label{eq:eigfun_d_singlet}
\end{equation}
where $|d_{\pm} \rangle$ are two degenerate $d$-wave-like eigenvectors chosen simultaneously as the eigenvectors of mirror symmetry $M$ with $M |d_{\pm} \rangle = \pm |d_{\pm} \rangle$. 
The second smallest eigenvalue of $V_{s}$ that fulfills the same 2D IR is $V_{s}^{p}$, characterized by 
\begin{equation}
V_{s} |p_{\pm} \rangle = V_{s}^{p}  |p_{\pm} \rangle,
\end{equation}
where $|p_{\pm} \rangle$ are two degenerate $p$-wave-like eigenvectors with $\pm$ defined similar to  $|d_{\pm} \rangle$. As discussed in the main text, the $d$-wave-like and $p$-wave-like states are distinguished by their dominant Fourier components.

Figure~\ref{fig:figureS9}(b) plots $\Pi^{pp}_{KK}(n)$, which possesses $C_3$ symmetry, indicating $V_{s} \Pi^{pp}$ is invariant under $C_3$ symmetry. For the limited case with $\Pi^{pp} = \chi_0 \mathcal{I}$ proportional to identify matrix $\mathcal{I}$, it follows directly that the eigenvectors of $V_s$ are also the eigenvectors of $V_{s} \Pi^{pp}$, and 
the critical temperature is given by $V_{s}^{d} \chi_0 =-1$. For the present study, we have
\begin{equation}
\Pi^{pp} = \chi_0 \mathcal{I} + \delta \Pi,
\end{equation}
where $\chi_0$ denotes the patch-averaged value of $\Pi^{pp}_{KK}(n)$ and $\delta \Pi$ represents the changes of particle-particle susceptibility around the annular Fermi surfaces. The eigenstates of $V_{s} \Pi^{pp} $ belonging to a particular IR can be expanded using the eigenstates of $V_s$ belonging to the same IR. By keeping the smallest and second smallest eigenvalues of $V_s$, the eigenvalues and eigenvectors of $V_{s} \Pi^{pp} $ can be obtained by solving the following $2\times 2$ matrix
\begin{equation}
(\langle p|, \langle d|) V_s\Pi^{pp} 
\begin{pmatrix}
|p \rangle \\
|d \rangle
\end{pmatrix} = 
\begin{bmatrix}
V_s^p ( \chi_0 +   \langle p| \delta \Pi |p \rangle )  &  V_s^{p}\langle p| \delta \Pi |d \rangle \\
V_s^{d} \langle d| \delta \Pi |p \rangle  & V_s^d (\chi_0  +  \langle d| \delta \Pi |d \rangle)
\end{bmatrix}
=\begin{pmatrix}
\epsilon_{p} & \delta_{p}  \\
\delta_{d} & \epsilon_{d}
\end{pmatrix},
\label{eq:eigen}
\end{equation}
where $|d \rangle$ and $|p \rangle$ denote either $|d_{+} \rangle$ and $|p_{+} \rangle$ or $|d_{-} \rangle$ and $|p_{-} \rangle$ because $\delta \Pi$ also possesses $M$ symmetry, and
\begin{equation}
\begin{aligned}
\epsilon_{p,d} &= V_s^{p,d} (\chi_0 + \langle p,d| \delta \Pi |p,d \rangle), \\
\delta_{p,d} &= V_s^{p,d}  \langle p,d| \delta \Pi |d,p \rangle.
\end{aligned}
\label{eq:2b2matrix}
\end{equation}
The above description resembles a two-level system, where the mixing between eigenvectors $|d \rangle $  and $|p \rangle $ is determined by the level spacing $|\epsilon_p-\epsilon_d|$ and inter-level coupling $\delta_{p,d} $.  Therefore, the small difference between the gap functions shown in Fig.~\ref{fig:figureS9}(a) may arises from two different sources. First, the variation of $\delta \Pi$ around the Fermi surface is small comparing to $\chi_0$. Secondly, the pairing interaction in the $d$-wave like channel $V_s^d$ is much smaller than other channels, as shown in Fig.~3(a) in the main text, leading to $\epsilon_d \ll \epsilon_p$. To further confirm these arguments, we evaluate the four entries of the $2\times 2$ matrix in Eq.~(\ref{eq:2b2matrix}) at the critical temperature $T_{c,d} \sim 1$ K. The results are listed in Table.~\ref{tab:table3}, where $\sqrt{\delta_d \delta_p} \ll \epsilon_p -\epsilon_d$ explains the negligible difference between the two results shown in Fig.~\ref{fig:figureS9}(a). 
Overall, we conclude that the structure of the gap function arises mainly from the momentum-space structure of the pairing interaction.

\begin{table}
\centering
\caption{Parameters in Eq.~(\ref{eq:eigen}) estimated at the critical temperature $T_{c,d}$ for the spin-singlet $d$-wave-like pairing channel.} 
\label{tab:table3} 
\begin{tabular}{ c | c |c ||c |c |c | c  }
\hline
\hline
$\chi_0$ (eV$^{-1}\cdot$nm$^{-2}$) & $V_s^{p} $ (eV$\cdot$nm$^{2}$) & $V_s^{d}$ (eV$\cdot$nm$^{2}$) & $\epsilon_p$ & $\epsilon_d$ & $\delta_p$ & $\delta_d$ \\
\hline
0.0704 & -5.1215 & -16.0625 & -0.3942 & -1.0025 & -0.0040 & -0.0124  \\
 \hline
 \hline
\end{tabular}
\end{table}

\subsection{Decomposition of the pairing interaction}
Here we present the details of the decomposition of the pairing interactions into the three types of scatterings illustrated in Fig.~3(e) in the main text. Taking the spin-singlet pairing channel as an example, the line labeled as “Total” in Fig. 3(f) in the main text plots the temperature dependence of the $d$-wave-like pairing interaction $V_{s}^{d}$, which is defined explicitly in Eq.~(\ref{eq:eigfun_d_singlet}). As shown in Fig.~2 in the main text, the spin-singlet pairing interaction $V_{s}$ possesses the following matrix structure
\begin{equation}
V_s = \begin{pmatrix}
V_{s,\text{inner}} & V_{s,\text{inter}} \\
V_{s,\text{inter}} & V_{s,\text{outer}}
\end{pmatrix},
\end{equation}
where $V_{s,\text{inner}}$ ($V_{s,\text{outer}}$) denotes scatterings within the inner (outer) annular Fermi surface,  $V_{s,\text{inter}}$ denotes scatterings between the inner and outer annular Fermi surfaces, as schematically illustrated in Fig.~3(e) in the main text. The lines labeled as “Inner”, “Inter”, and “Outer” in Fig.~3(f) are temperature dependence of $V_{s,\text{inner}}^{d}$, $V_{s,\text{inter}}^{d}$, and $V_{s,\text{outer}}^{d}$ calculated respectively by 
\begin{equation}
V_{s,\text{inner}}^{d} =  \langle d_{\pm} | 
\begin{pmatrix}
V_{s,\text{inner}} & 0 \\
0 & 0
\end{pmatrix}
|d_{\pm} \rangle,
\end{equation}
\begin{equation}
V_{s,\text{inter}}^{d} =  \langle d_{\pm} | 
\begin{pmatrix}
0& V_{s,\text{inter}}  \\
V_{s,\text{inter}}  & 0
\end{pmatrix}
|d_{\pm} \rangle,
\end{equation}
and
\begin{equation}
V_{s,\text{outer}}^{d} =  \langle d_{\pm} | 
\begin{pmatrix}
0& 0 \\
0 & V_{s,\text{outer}}
\end{pmatrix}
|d_{\pm} \rangle,
\end{equation}
where $|d_{\pm} \rangle$ are the doubly degenerate $d$-wave-like eigenvectors for the smallest eigenvalue $V_s^d$ of $V_s$ (see Eq.~(\ref{eq:eigfun_d_singlet})).
Therefore, we have the total pairing interaction in the $d$-wave-like channel $V_{s}^{d} = V_{s,\text{inner}}^{d} + V_{s,\text{inter}}^{d} + V_{s,\text{outer}}^{d}$. Based on these calculations, we are able to separate the contributions to the pairing interaction in particular channel from the three types of scatterings and single out the dominant one. As discussed in the main text, we find that the $d$-wave-like spin-singlet pairing is dominated by scatterings within the inner Fermi surface, while the $p$-wave-like spin-triplet pairing is dominated by scatterings between the inner and outer Fermi surfaces.

\section{Role of Exchange diagram}
In the main text, motivated by the observations that the particle-hole susceptibilities (Figs.~4(c)-(d) in the main text) associated with the EX diagrams possess similar momentum-space structures to those of pairing interactions $u_t(n_1,\bar{n}_1;\bar{n}_2,n_2)$ and $u_e(n_1,\bar{n}_1;n_2,\bar{n}_2)$ (Figs.~2(b)-(c) in the main text) at low temperatures, we argue that the momentum-space structures of $u_{t,e}$ are given by the following approximations:
\begin{equation}
\begin{aligned}
 \delta u_t &= u_t -\langle u_t \rangle  \propto -\langle u_t \rangle^2 \Pi^{ph}_{KK'} \\
\delta u_e  &= u_e -\langle u_e \rangle  \propto - \langle u_a \rangle \langle u_e \rangle \Pi^{ph}_{KK},
\end{aligned}
\end{equation}
where $\langle u_{a} \rangle$, $\langle u_{t} \rangle$, and $\langle u_{e} \rangle$ denote the averaged intra-valley, inter-valley, and valley-exchange interactions, respectively. Here we discuss in more detail about the validity of these approximations.

\begin{figure}
 \centering
\includegraphics[width=0.7\columnwidth]{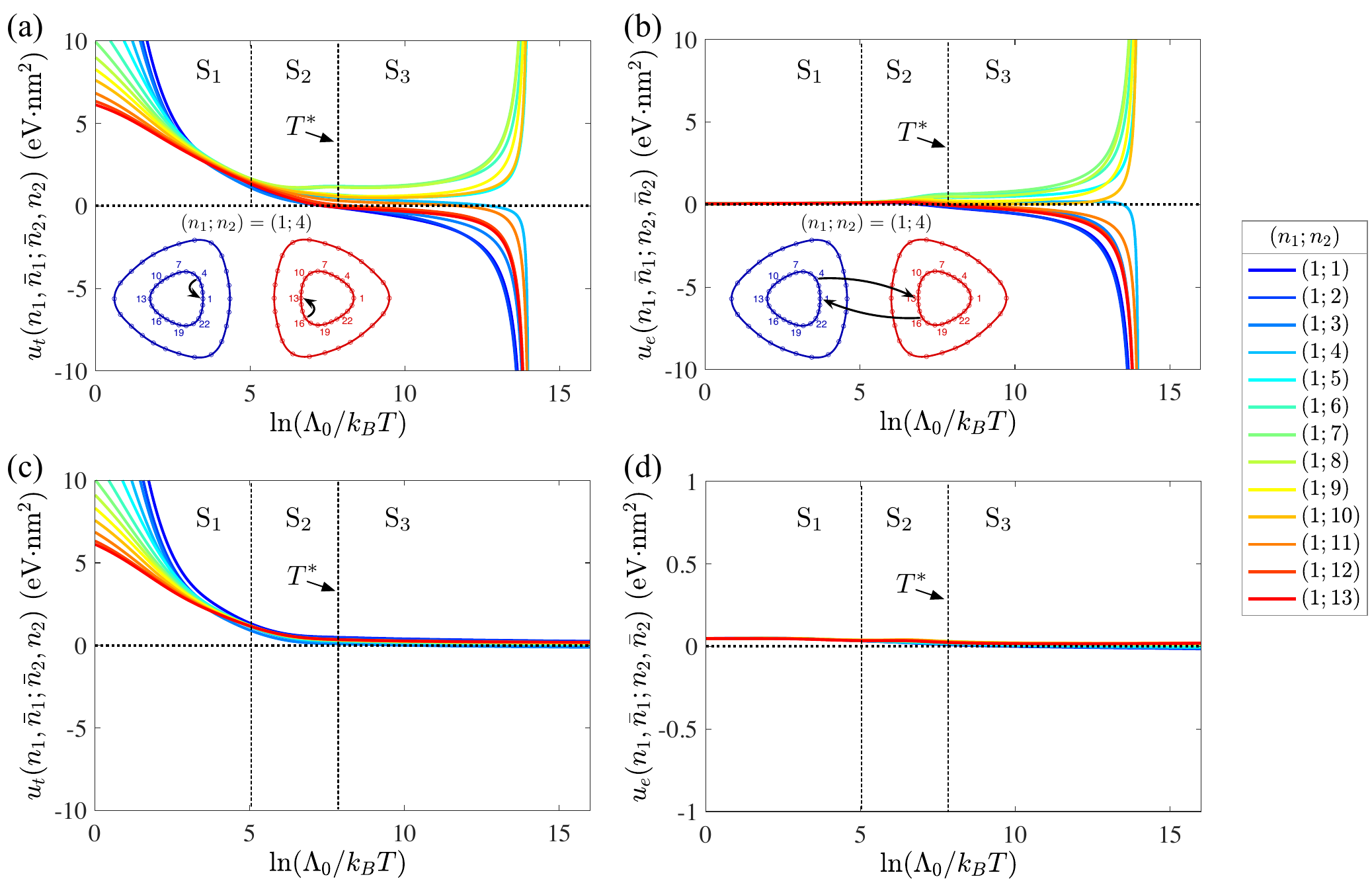}
 \caption{(a)-(b) Temperature flows of pairing interactions (a) $u_t(n_1,\bar{n}_1;\bar{n}_2,n_2)$ and (b) $u_e(n_1,\bar{n}_1;n_2,\bar{n}_2)$, where S$_{1,2,3}$ denote three stages of the RG flows, $T^* \sim 6$ K marks the lower temperature boundary of the S$_2$ stage. The inserts show the schematic diagrams of pairing interactions. (c)-(d) The temperature flows of (c) $u_t(n_1,\bar{n}_1;\bar{n}_2,n_2)$ and (d) $u_e(n_1,\bar{n}_1;n_2,\bar{n}_2)$ obtained by excluding the EX diagram (see Fig.~1(b) in the main text) when running the FRG calculations.  These calculations are performed by choosing $\Delta_d =30$ meV and $n_e =-1.75\times 10^{12}$ cm$^{-2}$.}
 \label{fig:figureS10}
\end{figure}

We first examine the role of the EX diagram by tracking the temperature flows of $u_{t,e}$. As shown in Figs.~\ref{fig:figureS10}(a)-(b), upon reducing temperature, $u_{t,e}$ gradually develop momentum-space structures that prefer $d$-wave-like pairing characterized by two sign changes of $u_{t,e}$ around the Fermi surface, consistent with the results shown in Figs.~2(b) and (c) in the main text. For comparison, Figs.~\ref{fig:figureS10}(c)-(d) plot temperature flows of $u_{t,e}$ obtained by excluding contributions from the EX diagram when running the FRG calculations. It is obvious that the EX diagram plays a crucial role in driving pairing instabilities at low temperatures. 

The temperature flows of $u_{t,e}$ can be viewed as having three stages marked by S$_{1,2,3}$ in Fig.~\ref{fig:figureS10}. The S$_1$ stage (high temperatures) is dominated by the famous particle-particle (PP) ladder diagram reduction of repulsive interaction strengths that is always important for superconductivity, and by the screening effects captured by forward scattering (FS). The EX enhancement becomes important in the S$_2$ stage (intermediate temperatures), giving rise to a particular momentum-space structures of  $u_{t,e}$. In the S3 stage (low temperatures), the particle-hole susceptibilities are nearly unchanged, while the zero-$q$ particle-particle susceptibility diverges logarithmically. Therefore, the S$_3$ stage is driven by the PP diagram, which amplifies the momentum-space structures developed in the S$_2$ stage, as indicated in Fig.~\ref{fig:figureS10}.

\begin{figure}
 \centering
\includegraphics[width=0.7\columnwidth]{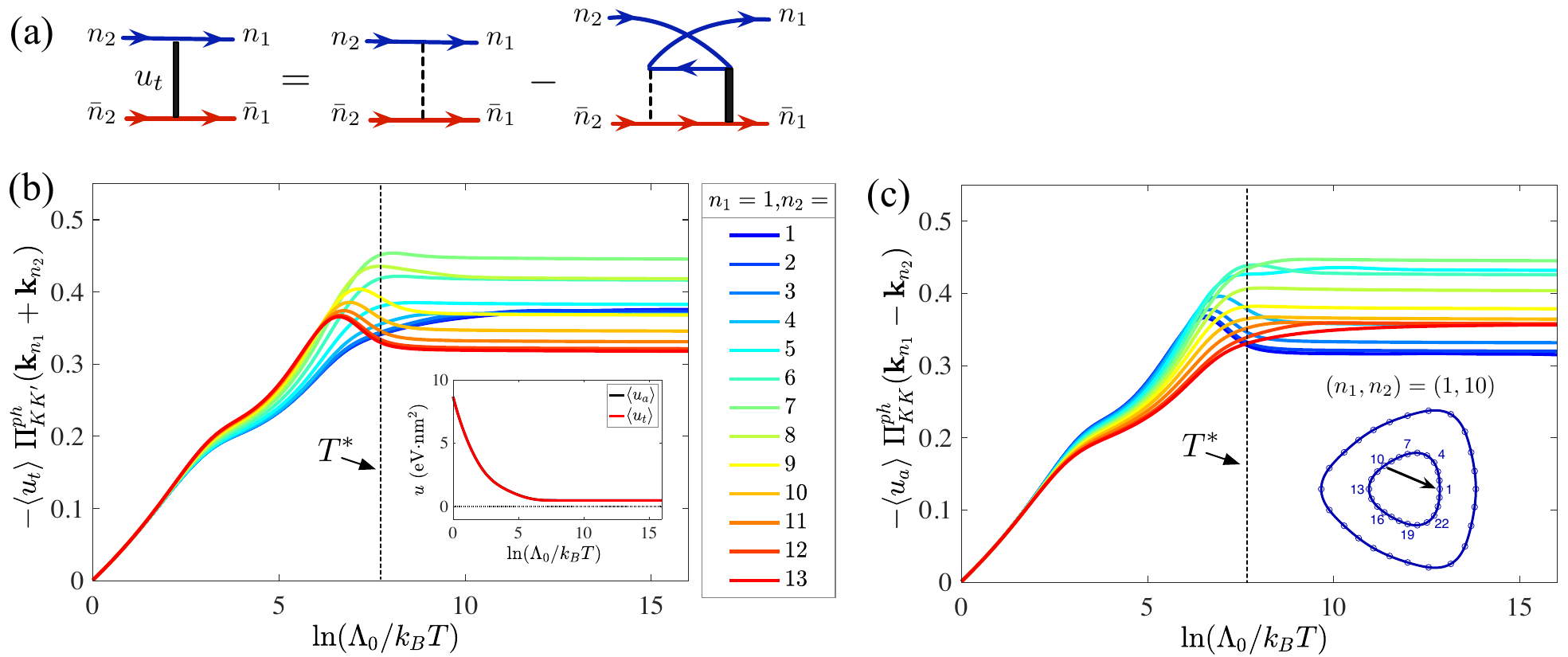}
 \caption{(a) Dyson expansion of the EX diagram for pairing interaction $u_t(n_1,\bar{n}_1;\bar{n}_2,n_2)$, where the dashed line denotes the renormalized vertex due to PP and FS diagrams, the solid rectangle represents the full vertex. (b) Temperature dependence of $- \langle u_t \rangle \Pi_{KK'}^{ph}$, where $\langle u_t \rangle$ (see insert) denotes the averaged inter-valley interaction and $ \Pi_{KK'}^{ph} $ is the inter-valley particle-hole susceptibility that associated with the EX diagram shown in (a). (c) Temperature dependence of $- \langle u_a \rangle \Pi_{KK}^{ph}$, where $\langle u_a \rangle$ denotes the averaged intra-valley interaction and $ \Pi_{KK}^{ph} $ is the intra-valley particle-hole susceptibility. The insert shows the definition of the transfer wavevector for $ \Pi_{KK}^{ph} (\bm{k}_{n_1}-\bm{k}_{n_2})$.}
 \label{fig:figureS11}
\end{figure}

To further examine the role of EX diagram in the S$_2$ stage, we next take $u_t$ as an example to evaluate the EX series expansion shown graphically in Fig.~\ref{fig:figureS11}(a) at temperature $T^*$ marked in Fig.~\ref{fig:figureS10}. The dashed line denotes the renormalized vertex due to PP and FS diagrams, and are given by the FRG flows of inter-valley interaction by excluding the contributions from the EX diagram, as illustrated partially in Fig.~\ref{fig:figureS10}(c). By comparing Figs.~\ref{fig:figureS10}(c) with (a), we notice that $u_t$ depends less on the scattering angle in Fig.~\ref{fig:figureS10}(c) at temperature $T^*$. Therefore, it is a good approximation to use the averaged value $\langle u_t \rangle$ to represent the dashed lines shown in Fig.~\ref{fig:figureS11}(a), leading to 
\begin{equation}
 u_t   \approx \frac{\langle u_t \rangle}{1+\langle u_t \rangle \Pi_{KK'}^{ph}},
\end{equation}
where $\Pi_{KK'}^{ph}$ is the inter-valley particle-hole susceptibility. As shown in Fig.~\ref{fig:figureS11}(b), the dimensionless parameters $-\langle u_t \rangle \Pi_{KK'}^{ph} \sim 0.4$. Therefore, we have
\begin{equation}
\delta u_t  = u_t-\langle u_t \rangle \propto  -\Pi_{KK'}^{ph} \langle u_t \rangle^2,
\end{equation}
which amounts to expanding the EX diagram up to second order. By comparing Fig.~\ref{fig:figureS11}(b) with Fig.~\ref{fig:figureS10}(a) at $T^*$, it is obvious that the momentum-space structure of $u_t$ is determined by that of $-\Pi_{KK'}^{ph}$, consistent with discussions in the fourth paragraph on page 3 of the main text. Similar arguments can be carried out for $u_e$, resulting in 
\begin{equation}
\delta u_e = u_e-\langle u_e \rangle \propto -\Pi_{KK}^{ph}  \langle u_e \rangle \langle u_a \rangle.
\end{equation}
The above approximation is valid because $-\langle u_a \rangle \Pi_{KK}^{ph} \sim 0.4$ as shown in Fig.~\ref{fig:figureS11}(c). Therefore, as we argued in the main text, the momentum-space structures of pairing interactions arise primarily from the momentum-space structures of the particle-hole susceptibilities associated with the EX diagrams

After the structure of $u_t$ is established in stage S$_2$, the RG flow equation for $u_t$ can be well captured by only including the PP diagram because the zero-$q$ particle-particle susceptibility ($\Pi_{KK}^{pp}$) diverges logarithmically at low temperatures, and the particle-hole susceptibilities are nearly temperature independent below $T^*$, as indicated in Fig.~\ref{fig:figureS11}(b). As depicted in Fig.~1(b) in the main text, the RG flow of pairing interaction $u_t$ therefore reads
\begin{equation}
\partial_{T} u_{t}(n_1;n_2) = -\partial_{T} \Pi_{KK}^{pp} \sum_{n}u_t(n_1;n)u_t(n;n_2),
\end{equation}
where, for simplification, only contributions from inter-valley interaction are included, $u_{t}(n_1;n_2)$ represents $u_{t}(n_1,\bar{n}_1;n_2,\bar{n}_2)$, and the patch-dependence of $\Pi_{KK}^{pp}$ is ignored. The above equation can be decoupled into RG flow equations for each eigenvalue of the matrix $u_{t}(n_1;n_2)$. For example, at $T^*$, $u_t(n_1;n_2)$ has a negative eigenvalue $\lambda_{d}(T^*)<0$ in the $d$-wave-like channel (symmetry of the corresponding eigenvectors). The above differential equation reduces to
\begin{equation}
\partial_{T} \lambda_d = -\partial_{T} \Pi_{KK}^{pp} \lambda_d^2,
\end{equation}
which has the solution
\begin{equation}
\lambda_d(T) = \frac{\lambda_d(T^*) }{1+\lambda_d(T^*) \delta \Pi^{pp}_{KK}(T) }
\end{equation}
where $\delta \Pi^{pp}_{KK}(T) = \Pi^{pp}_{KK}(T)-\Pi^{pp}_{KK}(T^*) \propto \ln (1/k_BT)$ diverges logarithmically, giving rise to a pole of $\lambda_d(T)$ at temperature $T \sim e^{1/\lambda_d(T^*)}$, as indicated in Fig.~\ref{fig:figureS10}(a). Therefore, the attractive interaction channels that develop in stage S$_2$ will be amplified by the PP diagram BCS theory process in stage S$_3$, while the repulsive channels will be suppressed logarithmically. 
 
 In the present work, the pairing instabilities occur in higher angular momentum e.g. $p$-, $d$-, $f$-wave channels, which are generated entirely from higher order terms of the pairing interaction. That is to say, for the channels we are examining, $\delta u_{t,e}$ are $u_{t,e}$. The averaged value ($s$-wave channel) of the pairing interaction does not contribute. We note that the full FRG calculations are of course more complex than our cartoon description of their essence, but we believe the central ingredients are captured by our narrative. 
 
 \section{Numerical discretization schemes }
In this work, there are two numerical discretization schemes. The first one is developed to approximate the 4PVs, namely, dividing the $\bm{k}$ space around each valley into 48 patches, as illustrated in Fig.~1(a) in the main text. The second scheme is developed for calculating the particle-particle and particle-hole susceptibilities by dividing the $\bm{k}$ space into much denser meshes because the susceptibilities are especially sensitive to the details of the band dispersion near Fermi energy at low temperatures. For RTG, the size of the annular Fermi surface is very small compared to the size of BZ. In order to balance accuracy and efficiency in calculating the particle-particle and particle-hole susceptibilities, we employ an improved discretization scheme by scaling the density of $\bm{k}$ points with the band energy $\epsilon(\bm{k})$. Specially, we choose higher $\bm{k}$-space resolution when the electron energy is closer to the Fermi level. For example, we choose 200000 random $\bm{k}$ points within the energy window $\epsilon \in[-0.002, +0.002]$ meV. Within this improved scheme, we have tested that the temperature derivatives of the particle-particle and particle-hole susceptibilities are sufficiently smooth down to $T=1$ mK. For temperature-flow FRG, the inherent thermal broadening of the Fermi-Dirac distribution function enables us to choose a relative lower $\bm{k}$-space resolution when band energy is away from the Fermi level.
  
\begin{figure}
 \centering
\includegraphics[width=0.99\columnwidth]{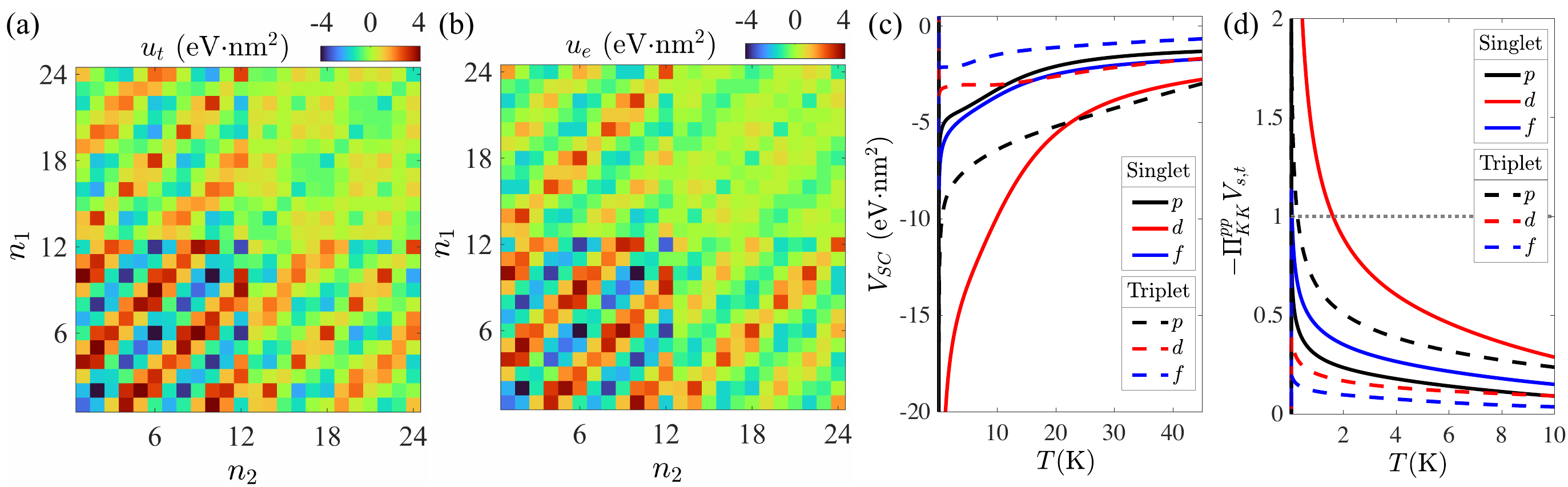}
 \caption{24-patch scheme. (a)-(b) Renormalized pairing interactions (a) $u_t(n_1,\bar{n}_1;\bar{n}_2)$ and (b) $u_e(n_1,\bar{n}_1;n_2)$ at temperature $T=70$ mK. (c) Temperature flows of pairing interaction in different channels. (d) Temperature dependence of the eigenvalues of $-\Pi^{pp}_{KK}V_{s,t}$, where $T_c's$ are estimated by the temperatures of intersections between these eigenvalues and 1 (dotted line). These calculations are obtained by choosing $\epsilon=5$, $d_s=40$ nm, $\Delta_d = 30$ meV, and $n_e = -1.75\times10^{12} $ cm$^{-2}$.}
 \label{fig:figureR2}
\end{figure}

\begin{figure}
 \centering
\includegraphics[width=0.99\columnwidth]{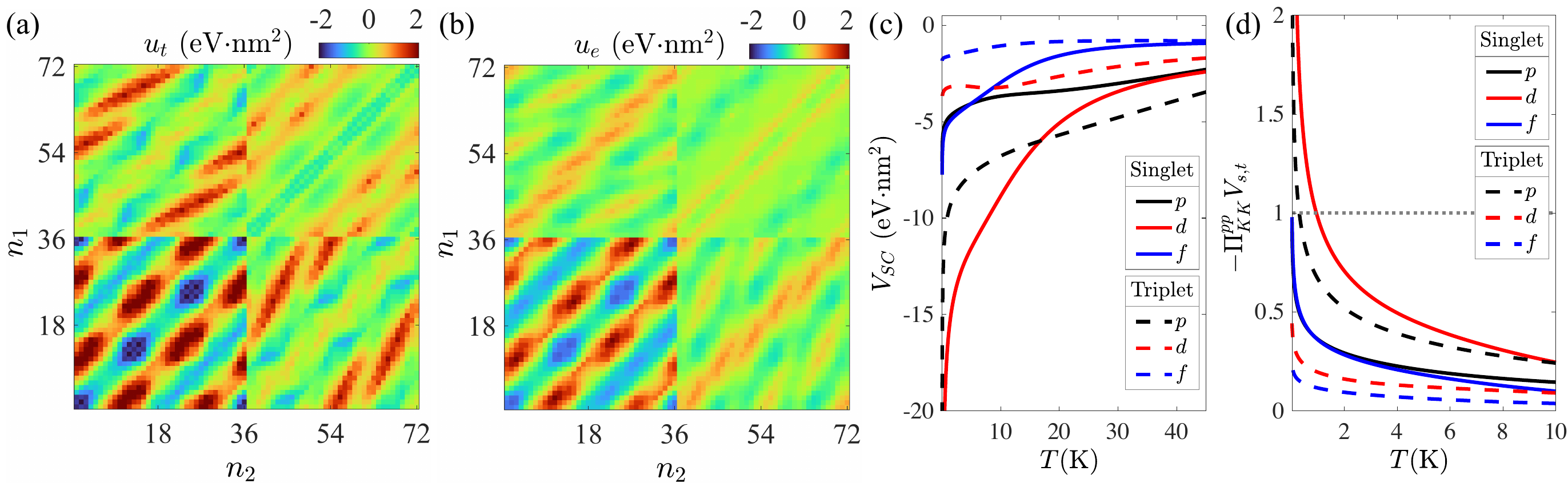}
 \caption{72-patch scheme. (a)-(b) Renormalized pairing interactions (a) $u_t(n_1,\bar{n}_1;\bar{n}_2)$ and (b) $u_e(n_1,\bar{n}_1;n_2)$ at temperature $T=70$ mK. (c) Temperature flows of pairing interaction in different channels. (d) Temperature dependence of the eigenvalues of $-\Pi^{pp}_{KK}V_{s,t}$, where $T_c's$ are estimated by the temperatures of intersections between these eigenvalues and 1 (dotted line). These calculations are obtained by choosing $\epsilon=5$, $d_s=40$ nm, $\Delta_d = 30$ meV, and $n_e = -1.75\times10^{12} $ cm$^{-2}$.}
 \label{fig:figureR3}
\end{figure}

By keeping the $\bm{k}$-space discretization scheme for computing the susceptibilities unchanged, we run the FRG calculations using both the 24-patch and 72-patch schemes, with the results plotted in Fig.~\ref{fig:figureR2} and Fig.~\ref{fig:figureR3}, respectively. By comparing Fig.~\ref{fig:figureR2}(a)-(b) and Fig.~\ref{fig:figureR3} (a)-(b) with Fig.~2(b)-(c) in the main text, we find that the momentum-space structures of the pairing interactions exhibit similar features in all the three patch schemes, preferring a $d$-wave-like character on the inner annular Fermi surface. The effective interaction strengths at $T=70$ mK calculated from the 24-patch scheme are larger than those calculated from the 48- and 72-patch schemes. By comparing Fig.~\ref{fig:figureR2}(c) and Fig.~\ref{fig:figureR3}(c) with Fig~3(a) in the main text, we find that the leading instability, namely, the spin-singlet $d$-wave-like pairing, keeps unchanged for all the three patch schemes. Table \ref{tab:tableS1} lists the $T_c's$ for leading pairing channel, where a saturation has been reached at the 48-patch scheme, justifying the validity of the patch scheme adopted in the main text. Overall, we conclude that a spare patch discretization is more likely to influence the quantitative values of $T_c$ estimation than to alter the qualitative features of the leading instability, and that the 48-patch scheme employed in the main text is sufficient for the present study.

\begin{table}[ht]
\centering
\caption{Critical temperatures $T_{c,d}$ of $d$-wave pairing channel estimated from three different patch schemes.}
\label{tab:tableS1} 
\begin{tabular}{ c | c |c |c   }
\hline
\hline
 Patch scheme & 24-patch & 48-patch & 72-patch \\
  \hline
 $T_{c,d}$ & 1.61 K  &  0.92 K & 0.94 K \\
 \hline
 \hline
\end{tabular}
\end{table}